\documentclass{article}
\usepackage{arxiv}
\usepackage{textcomp}
\usepackage[T1]{fontenc, url}
\usepackage[utf8]{inputenc}
\usepackage{titlesec}
\usepackage{multirow}
\usepackage{adjustbox}
\usepackage{graphicx}
\usepackage{amsmath, amssymb, amsthm} 
\usepackage{parskip} 
\usepackage{color}
\usepackage{subcaption} 
\usepackage{appendix,setspace}
\usepackage{hyperref}
\usepackage{chngcntr}
\counterwithin{table}{section} 
\counterwithin{figure}{section} 
\numberwithin{equation}{section} 
\newtheorem{theorem}{Theorem}[section]
\newtheorem{lemma}[theorem]{Lemma}
\newtheorem{corollary}[theorem]{Corollary}
\newtheorem{definition}{Definition}[section]
\usepackage{enumitem} 
\newlist{condenum}{enumerate}{1} 
\usepackage{xparse,nameref}
\usepackage[bottom]{footmisc} 

\usepackage[english]{babel} 
\graphicspath{{Images/}{../Images/}} 
\usepackage{microtype}

\titleformat*{\section}{\LARGE\bfseries} 
\titleformat*{\subsection}{\Large\bfseries} 
\titleformat*{\subsubsection}{\large\bfseries} 

\usepackage{fancyhdr}
\usepackage{subfiles}
\usepackage[rightcaption]{sidecap}
\usepackage{wrapfig}
\usepackage{float}
\usepackage[labelfont=bf]{caption} 
\usepackage[para]{threeparttable} 
\usepackage{url}
\usepackage[table,xcdraw]{xcolor}
\usepackage{physics}
\usepackage{graphicx}
\usepackage{natbib}

\title{Robust and Efficient Parameter Estimation for Discretely Observed Stochastic Processes}

\author{
   Rohan Hore\\
   Physical Sciences Division\\
   University of Chicago\\
   Chicago,United States\\
  \texttt{rohanhere@uchicago.edu} \\
  \And
   Abhik Ghosh\\
  Interdisciplinary Statistical Research Unit\\
  Indian Statistical Institute,Kolkata\\
  WestBengal,India \\
  \texttt{abhik.ghosh@isical.ac.in} \\
}

\begin{document}
\maketitle
\def\biblio{} 
\begin{abstract}
    
In various practical situations, we encounter data from stochastic processes which can be efficiently modelled by an appropriate parametric model for subsequent statistical analyses. 
Unfortunately, the most common estimation and inference methods based on the maximum likelihood (ML) principle are susceptible to minor deviations from assumed model or data contamination 
due to their well known lack of robustness. Since the alternative non-parametric procedures often lose significant efficiency, in this paper, 
we develop a robust parameter estimation procedure for discretely observed data from a parametric stochastic process model which exploits the nice properties of 
the popular density power divergence measure in the framework of minimum distance inference. 
In particular, here we define the minimum density power divergence estimators (MDPDE) for the independent increment and the Markov processes. 
We  establish the asymptotic consistency and distributional results for the proposed MDPDEs in these dependent stochastic process set-ups 
and illustrate their benefits over the usual ML estimator for common examples like Poisson process, drifted Brownian motion and auto-regressive models. 

\par\vspace*{\fill} 
\textbf{\textit{Keywords --}} Stochastic Process; Robust Estimation; Density Power Divergence;  Poisson Process; Drifted Brownian Motion; Autoregressive Process. 

\biblio 

\end{abstract}
\section{Introduction} 

The basic setup of statistical inference assumes the availability of a simple random sample from the distribution of interest
and one needs to perform parameter estimation and/or tests of hypothesis based on such 
independent and identically distributed (IID) sample observations. 
However, in real life we do not always get IID observations; rather, we often may encounter a dependent set of observations, 
where the associated random variables together come from a joint distribution. 
The dependence structure of the associated random variables can be very arbitrary, depending on the underlying data generating process. 
Accordingly, the problem of statistical inference in this case can be quite non-routine 
and hence useful methods dealing with such situations are of great practical value. 
In this paper, we will deal with (dependent) observations from a stochastic process having density as defined below.

\begin{definition}[Stochastic Process with density] 
A collection $\{X(t)\}_{t\geq 0}$ of random variables is called a stochastic process having density if,  for any k and $0\leq t_1 \leq t_2 \leq \hdots \leq t_k<\infty$,
the random vector $(X(t_1),X(t_2),\hdots,X(t_k))$ has a density $g(\cdot;t_1,t_2,\hdots,t_k)$ with respect to some common dominating measure. 
\end{definition}

For efficient inference about a stochastic process, 
we need to model the densities $g(\cdot; t_1, t_2, \hdots, t_k)$ with a parametric  family  of densities 
of the form $f(\cdot; \boldsymbol{\theta}, t_1, t_2, \hdots, t_k)$, where $\boldsymbol{\theta}$ is a (vector) parameter to be estimated based on the observed data. 
It is often expensive to observe the whole stochastic process for a continuous interval of time. 
Hence, one rather records observations for a prefixed (discrete and finite) time-stamp vector 
$\boldsymbol{t}=(t_0=0, t_1, \ldots, t_n)$ for some $n\geq 1$, 
so that the corresponding sample realization looks like $\{X(t_i)\}_{0\leq i \leq n}$. 
Hence, practically speaking, we basically record finitely many (dependent) observations from a continuous-time stochastic process. 
Then, our fundamental task will be to estimate the unknown model parameters based on these observations $\{X(t_i)\}_{0\leq i\leq n}$. 
Since the dependence structure can be very arbitrary, we will start with the independent increment processes 
and subsequently extend our ideas to Markov processes. 
Through out the paper, we will consider our time-stamp vector $\boldsymbol{t}= \{t_i\}_{i\geq 0}$ to be fixed before the data collection.

Finding an efficient estimator for such parametric stochastic process models is not a new topic.  
Some notable works towards establishing an estimator for parametric stochastic processes can be found long back in \cite{borgan1984maximum}, \cite{ryden1996algorithm}, 
\cite{hossain1993estimating}, \cite{zhao1996maximum} as well as more  recently in  \cite{tanaka2008parameter}. 
The most natural estimator is the maximum likelihood estimator (MLE) which is easy to compute 
and can be shown to be asymptotically normal having maximum possible efficiency under suitable regularity conditions.
Such nice properties of the MLE are known even for dependent observations; see, e.g., \cite{bhat1974method} and \cite{prasad1976maximum}. 
But, the MLE does struggle if the underlying data is contaminated. 
This lack of robustness has been observed in the literature for different known models, e.g., 
in  \cite{xu2011robustness}, \cite{weems2004robustness} or \cite{genton1999robustness}. 
Here, we start with an example of Poisson process to show that the classical MLE fails 
when there is even a minute contamination in the data. 


\subsection{Motivation: Robustness Issue of the MLE}
\label{SEC:Motivation}

Let us consider the familiar setup of a general Poisson Process as defined below. 

\begin{definition}[Poisson Process]
A stochastic process $\{N(t)\}_{t\geq 0}$ is called a Poisson process if it has independent increments, i.e., 
for any given $n$ and $0\leq s_1\leq s_2 \hdots \leq s_n$, the increments $N(s_i)-N(s_{i-1})$ are independent across $i= 1, 2 \hdots, n$,
$N(0)=0$ and $N(t)\sim Poi(\Lambda(t,\theta))$ for all $t>0$, with $\theta\in \mathbb{R}$
and $\Lambda(t,\theta)=\int_0^{t} \lambda(s,\theta)ds$; we call $\lambda(\cdot,\theta)$ the intensity function of the process.
\label{DEF:Poisson_proc}
\end{definition}

\begin{figure}[!h]
        \centering
        \includegraphics[scale=0.7]{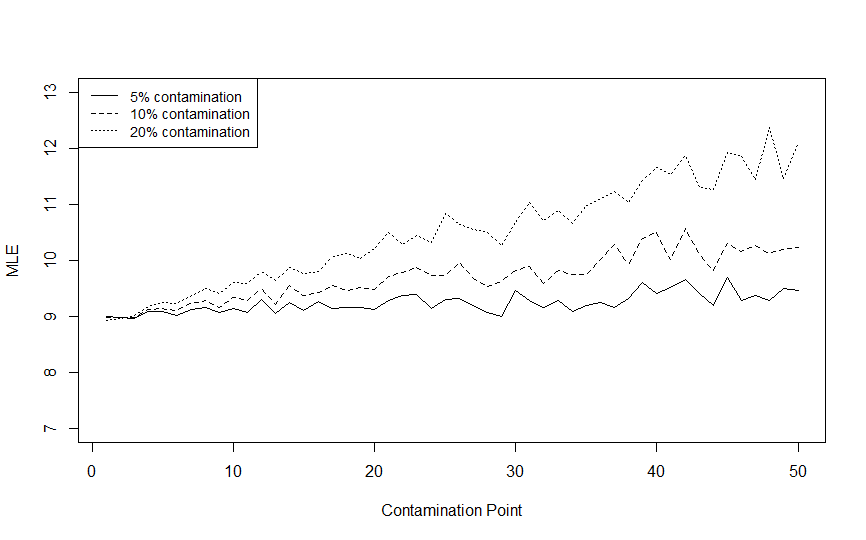}
        \caption{Effect of data contamination on the MLE for a Poisson Process}
        \label{contamination effect on mle}
    \end{figure}

Now for illustration purpose, let us prefix a time-stamp vector as $\boldsymbol{t}=\mathcal{Z}^{+}\cup\{0\}$
and consider discretely observed data from a Poisson process with intensity function $\lambda(t,\theta)=\frac{\theta}{2\sqrt{t}}$, $\theta \in \mathbb{R}$
and the integrated intensity function $\Lambda(t,\theta)= \theta \sqrt{t}$. 
Here our objective is to estimate $\theta$. We simulate sample data of size $n=25$ with the true parameter value $\theta^g=9$ and 
contamination index 5,  i.e., we will only contaminate the $5$-th increment distribution and other increment distributions remain unchanged. 
Say, our contaminated $5$-th increment distribution is given by
\begin{equation*}
    (1-p_{cont})*Poi(\Lambda(t_5;\theta)-\Lambda(t_4,\theta))+p_{cont}*\Lambda_{r},
\end{equation*}
where $r$ is the contamination point and $p_{cont}$ is the contamination proportion. 
The effects of such contamination on the MLE, for varying $r$ and $p_{cont}$, are shown in the Figure \ref{contamination effect on mle}.
Clearly, we can see from the above example that, as we take our contamination point and contamination proportion (percentage) higher, the bias of MLE also increases. This clearly depicts high sensitivity of the MLE towards outliers. 
Thus, under presence of contamination in the data, MLE is unable to perform well 
and any inference based on MLE, for example using asymptotic distributional results of the MLE, might be misleading.
So, it is an important task to find a suitable robust and efficient estimator of the model parameters for discretely observed 
parametric stochastic processes. 

\subsection{Literature Survey}

As per our knowledge, as of now, not much work and development has been done on finding robust estimators 
for general stochastic processes except for some scattered attempt under particular model assumptions which we briefly summarize here. 

\cite{kulkarni1987optimal} described a general method of constructing robust quasi-likelihood estimating functions 
for discrete time stochastic processes. In \cite{yoshida1990robust}, robust M-estimators were proposed for 
the particular example of Poisson process with periodic intensities and their asymptotic properties were investigated. 
The robustness issue of the estimation problem for in-homogeneous Poisson processes has been addressed in \cite{assunccao1999robustness},
based on the corresponding developments for the general Point process in \cite{assuncao1995robustness}. 
In \cite{stoimenova2005robust}, a robust efficient estimator has been proposed for the special example of branching processes 
with power series offspring distribution. Similar work has also been done in \cite{stoimenova2011robust}. 
In \cite{fay2007estimation}, a robust estimator of the memory parameter of an infinite source Poisson process, a particular case of $M/G/\infty$ queue process, has been proposed. 
Some other papers worth mentioning are \cite{moklyachuk2006robust} and \cite{moklyachuk2015minimax} 
which talks about the general robust estimation problems faced in stochastic processes. 
In \cite{rieder2012robust} optimally robust asymptotic linear estimates for Ornstein-Uhlenbeck process has been proposed. 
\cite{wang2013robust} have described robust estimators for the Weibull process. 
Some progresses in finding efficient robust estimator under the class of  diffusion processes have been done in \cite{yoshida1988robust},
following the M-estimation approach, and in \cite{lee2013minimum}, \cite{song2017robust}, \cite{song2020robust} and \cite{song2007minimum}
following the minimum density power divergence (DPD) approach. 
Recently, \cite{amoako2020robust} have noted down some robust statistical procedures 
for parametric estimation under the  geometric Brownian motion with application to stock price prediction.

Besides the above mentioned works, there has been a branch of literature discussing the robust inference for a particular
subclass of stochastic processes, namely the time series models. 
\cite{denby1979robust} described a class of M-estimators for estimating the parameter of a first order auto-regressive time series. 
\cite{chang1988estimation} proposed a procedure for estimating auto-regressive parameters in AR(1) model addressing
two special cases of outliers, namely innovational and additive outliers. 
In \cite{muler2009robust}, a class of M estimators have been proposed for the general ARMA model. 
Other recent works in this branch include \cite{kim2013robust}, \cite{kim2017robust} and \cite{park2017robust}. 
Specifically in the setup of Poisson autoregressive models, there are a few recent developments as reported in \cite{kang2014minimum}, \cite{kang2020robust} and \cite{kim2017robust}.
However, all these proposals are mostly confined to the special structures of the time series models and, hence,
do not directly translates to the cases of more general stochastic processes.

In a very recent work, \cite{Ghosh:2021} proposed robust parametric inference procedures for the finite Markov chain models
with general transition probability structures using the minimum DPD approach.
The DPD measure and its use in robust estimation were initially introduced in \cite{basu1998robust} for IID set-ups,
which subsequently becomes very popular due to its several nice properties. 
These includes high asymptotic efficiency along with the desired level of robustness, relatively easier computation and
an intuitive explanation as a robust generalization of the MLE; see Section \ref{SEC:MDPDE} for a brief review. 
For these reasons the minimum DPD estimator (MDPDE) has been later  extended to several complex model examples 
\citep[see, e.g.,][among many others]{Basu/etc:2011,ghosh2013robust,ghosh2016robust, ghosh2019robust,kim2020robust}. 
A general extension of its theory under the independent non-homogeneous observations has been developed in \cite{ghosh2013robust}.
In this paper, we will extend this idea to propose robust parameter estimates for more general dependent data structures 
arising from some important classes of stochastic process models.

\subsection{Our Contributions}


Particularly, in this paper, we will consider the problem of robust parameter estimation for discretely observed data 
from two broad classes of stochastic processes, namely the independent increment processes (IIPs) and the Markov processes (MPs),
defined as follows. 

\begin{definition}[Independent Increment Process]
An independent increment process (IIP) is defined as a continuous time stochastic process with following structure: 
	\begin{align*}
	X(t_i)- X(t_{i-1})   & \text{  are independent across $i= 1, 2, \ldots, n$},
	\\& \text{ for any given $n\geq 1$ and } 0=t_0\leq t_1 \leq t_2 \leq \ldots \leq t_n.
	\end{align*}
\end{definition}

\begin{definition}[Markov Process]
A Markov process (MP) is defined as a continuous time stochastic process with following structure: 
	\begin{align*}
	X(t_i)|X(t_{i-1}),X(t_{i-2}),& \ldots,X(t_1) \overset{d}{=} X(t_i)|X(t_{i-1}), 
\text{ for any given $n\geq 1$ and } 0=t_0\leq t_1 \leq \ldots \leq t_n.
	\end{align*}
\end{definition}

This paper develops the robust parameter estimation procedure by extending the definition of the MDPDE and deriving its properties
for discretely observed data from these two classes of stochastic processes, namely the IIPs and the MPs.
Under the IIP set-ups, the densities of the incremental distributions (which are independent but non-homogeneous) 
are used for parameter modeling  and subsequent definition of the MDPDEs, 
whereas the conditional densities are used for these purposes under the MP set-ups.
Although the MDPDEs are there in the literature for some time now and popularly applied to different set-ups,
they had never been studied for the general dependent set-ups of IIPs and MPs before our present attempt. 

Further, although extending the definition of the MDPDEs may seem somewhat intuitive,
the study of their theoretical properties under the dependence structures of the IIPs and the MPs require non-trivial 
generalization of the existing theory and necessary assumption, which are the core contributions of the present paper. 
Besides developing the general set of sufficient conditions for consistency and asymptotic normality of our newly defined MDPDEs
for the general IIPs and MPs, we further examine their implications and validity for three important subclasses, 
namely the single family of increment distributions and the location-scale family of increment distributions in the IIPs 
and the m-dependent stationary Markov processes under MPs. We simplify the sufficient conditions and the asymptotic variances 
of the MDPDEs for these special cases illustrating the importance of our general theory, 
which can similarly be applied to other relevant subclasses of stochastic process models in a similar way.   
Additionally, a theoretical robustness study of the MDPDEs under the IIP models, via the classical influence function analyses, is presented in an Appendix to the paper.
For brevity in presentations, the proofs of most theoretical results (and the associated lemmas) are also moved to the Appendices.

Subsequently, we also numerically illustrate the performances of the MDPDEs, 
with specific focuses on the claimed robustness under data contamination,  
for important practical examples of Poisson process, drifted Brownian motions and autoregressive ()AR) process. 
The results support our theoretical derivations in terms of asymptotic efficiency.
The robustness of the MDPDEs under the IIPs is also supported theoretically via the influence function analysis from the existing literature of robust inference. 
The MDPDEs, as a solution of certain convex optimization problems, can be easily computed 
and performs quite competitively in terms of computation time with the other usual estimators;
efficient \textbf{\texttt{R}} codes are developed for the computations of the MDPDEs in our numerical illustrations
which are available from the authors on request.  
The major improvement of these estimators over existing MLE are in terms of robustness under high data contamination, 
without losing significantly in efficiency under pure data.



\biblio 
\section{The Minimum DPD Estimator for Stochastic Process Models}
\label{SEC:MDPDE}

Let us start with a brief background of the MDPDE for the sake of completeness.  
The DPD is a measure of discrepancy between two probability density functions $g$ and $f$, 
which is defined in terms of a single tuning parameter $\alpha \geq 0$ as \citep{basu1998robust}
\begin{equation}\label{Density Power Divergence Definition}
    d_\alpha(g,f)=\begin{cases}
    \int \{f^{1+\alpha}-(1+\frac{1}{\alpha})f^\alpha g+\frac{1}{\alpha}g^{1+\alpha}\}, &\text{if $\alpha > 0$,}\\
    \int g \ln(\frac{g}{f}), & \text{if $\alpha=0$},
    \end{cases}
\end{equation}
where $\ln$ represents the natural logarithm. 
While the divergence is not defined for $\alpha = 0$, $d_0(\cdot,\cdot)$ represents the divergence obtained in the limit as $\alpha \to 0$,
and is a version of the Kullback-Leibler divergence (KLD). 
On the other,  hand $\alpha = 1$ generates the squared $L_2$ distance.
When we have IID observations $Y_1, \ldots, Y_n$ from a popuation having density $g$
and we model them by a parametric family of densities 
$\mathcal{F}$ = $\{f(y;\boldsymbol\theta):\boldsymbol\theta \in \Theta \subseteq \mathbb{R}^p\}$,
then the MDPDE of the unknown parameter $\boldsymbol{\theta}$ is defined as the minimizer of 
$d_\alpha(\widehat{g}_n,f(\cdot;\boldsymbol\theta))$ over $\boldsymbol\theta \in \Theta$,
where $\widehat{g}_n$ is an empirical estimate of $g$ based on the observed sample $Y_1, \ldots, Y_n$.
A major advantage of the form of the DPD is that we do not need to use complex non-parametric smoothing (e.g., kernel estimators) 
to get an workable $\widehat{g}$; the third term in the form of the DPD is independent of $\boldsymbol{\theta}$ (so it can be ignored)
and the second term can be estimated by rewriting it as $(1+\frac{1}{\alpha})\int f(y;\boldsymbol\theta)^\alpha dG(y)$ 
and just using the empirical distribution function in place of $G$, leading to the simpler objective function given by 
\begin{equation*}
H_{n, \alpha}(\boldsymbol\theta)=\frac{1}{n}\sum_{i=1}^n\Bigg[\int f^{1+\alpha}(y;\boldsymbol\theta)dy
-\left(1+\frac{1}{\alpha}\right)f^{\alpha}(Y_i;\boldsymbol\theta)\Bigg].
\end{equation*}

The above framework of the MDPDE has been extended for the independent but non-homogeneous (INH) observations by \cite{ghosh2013robust}. 
If the observed data $Y_1,Y_2,\ldots,Y_n$ are independent but, for each $i$, $Y_i \sim g_i$ 
with $g_1,g_2,\ldots,g_n$ being possibly different densities with respect to some common dominating measure,
we can model them by the families of densities $\mathcal{F}_{i}$ = $\{f_i(\cdot;\boldsymbol\theta)|\boldsymbol\theta \in \Theta \}$,
respectively, for all $i=1, \ldots, n$, all sharing the same parameter $\boldsymbol\theta$. 
Then, \cite{ghosh2013robust} proposed to define the MDPDE of $\boldsymbol{\theta}$ 
by minimizing the average DPD measures between the sample observations and the model densities, i.e.,
$\frac{1}{n}\sum_{i=1}^n d_{\alpha}(\widehat{g}_i,f_i(\cdot;\boldsymbol\theta))$,
where $\widehat{g}_i$ is density of the degenerate distribution at $Y_i$. 
By a similar argument as above, the MDPDE can equivalently be defined by the minimizer of a simpler objective function given by 
\begin{equation*}
H_{n, \alpha}(\boldsymbol\theta)=\frac{1}{n}\sum_{i=1}^n\Bigg[\int f_i^{1+\alpha}(y;\boldsymbol\theta)dy-\left(1+\frac{1}{\alpha}\right)f_i^{\alpha}(Y_i;\boldsymbol\theta)\Bigg].
\end{equation*}
The asymptotic and robustness properties of the resulting MDPDEs are derived under appropriate conditions 
for the IID data in \cite{basu1998robust,Basu/etc:2011} and for the INH data in \cite{ghosh2013robust}.

\subsection{Parametric Modeling and Estimation for the Independent Increment Processes}
\label{SEC:MDPDE_IIP}

Let us consider discretely observed data $X_0:=0, X_1, \ldots, X_n$ from an IIP $\{X(t)\}_{t\geq 0}$ 
as the realizations of $\{X(t_i)\}_{0\leq i \leq n}$ for some $n\geq 1$ and time-stamp vector $\boldsymbol{t}$. 
Now, observe that the realization of $\{X(t_i)\}_{0\leq i \leq n}$ is equivalent to the realization of 
$\{X(t_i)- X(t_{i-1})\}_{1\leq i \leq n}$.
So, denoting $Y_i:= X_i-X_{i-1}$ for $i=1, \ldots, n$, we get that $Y_1, Y_2, \ldots, Y_n$ are indeed INH observations, 
since the original realizations $X_0, X_1, \ldots, X_n$ come from an IIP. 
Thus, we can readily use the formulations of the MDPDE from \cite{ghosh2013robust} 
by assuming an appropriate parametric model for $Y_i$s based on  that of the underlying IIP.  
To this end, in order to incorporate appropriate parametric assumptions,  
we first reformulate the IIP $\{X(t)\}_{t\geq 0}$  as follows:
\begin{align*}
   & X(t,s):= X(t)-X(s); X(0)=0 \text{ with } X(t,s)\sim g(\cdot;t,s) \text{ for any $t>s$}\\
   & X(t_i,t_{i-1})   \text{  are independent across $i$} \text{ for any given $n$ and } 0=t_0\leq t_1 \leq t_2 \leq \ldots \leq t_n .
\end{align*}
Since the time-stamp vector is prefixed as $\boldsymbol{t}= \{t_i\}_{i\geq 0}$, we will only observe 
$X(t_i,t_{i-1}):=X(t_i)- X(t_{i-1})$ having (increment) density $g(\cdot; t_i,t_{i-1})$ for $i=1, \ldots, n$. 
We want to model $g(\cdot;t_i,t_{i-1})$ by a parametric family 
$\mathcal{F}_{i,\boldsymbol{t}}$ = $\{f(\cdot;\boldsymbol{\theta},t_i,t_{i-1})|\boldsymbol{\theta} \in \Theta\subseteq \mathbb{R}^p \}$, 
respectively, for each $i= 1, 2, \ldots, n$. Observe that, given $t_i$,  
$f(\cdot;\boldsymbol{\theta},t_i,t_{i-1})$ can be very different across $i$ 
but they all share the same (vector) parameter $\boldsymbol{\theta}$ which we want to estimate. 
Thus, with above formulation, $\{Y_i:=X_i - X_{i-1}\}_{1\leq i \leq n}$ is a set of INH observations. 
So, following the idea of \cite{ghosh2013robust}, the MDPDE  of $\boldsymbol{\theta}$ can be defined as 
a minimizer of the average discrepancy between the data points and the model, measured in terms of the DPD, 
given by
\begin{equation*}
    \frac{1}{n}\sum_{i=1}^n d_\alpha(\widehat{g}_i,f_i(\cdot;\boldsymbol{\theta} ,t_i,t_{i-1})),
\end{equation*}
where $\widehat{g}_i$ is a density estimate for $Y_i$. In particular, we can take the dirac measure at observed $Y_i$  as a $\widehat{g}_i$ and,
as before, the MDPDE of $\boldsymbol{\theta}$ can equivalently be obtained by minimizing the simpler objective function 
\begin{equation}\label{Independent Increment Objective function}
H_{n, \alpha}(\boldsymbol{\theta};\boldsymbol{t})
= \frac{1}{n}\sum_{i=1}^n\Bigg[\int f^{1+\alpha}(y;\boldsymbol{\theta},t_i,t_{i-1})dy
-\left(1+\frac{1}{\alpha}\right)f^{\alpha}(Y_i;\boldsymbol{\theta},t_i,t_{i-1}) \Bigg]
=\frac{1}{n}\sum_{i=1}^n V_{\alpha}^I(Y_i|\boldsymbol{\theta},t_i, t_{i-1}),
\end{equation}
where $V_{\alpha}^I(Y_i|\boldsymbol{\theta},t_i, t_{i-1})$ indicates the term within square brackets
(superscript $I$ indicates that the underlying process is IIP). 
Upon required differentiation and simplifications, we get the estimating equation for the MDPDE as given by 
\begin{equation}\label{Independent Increment Estimating Equation}
    \sum_{i=1}^n\bigg[f^{\alpha}(Y_i;\boldsymbol{\theta},t_i,t_{i-1})\boldsymbol{u}(Y_i;\boldsymbol{\theta},t_i,t_{i-1})-\int f^{1+\alpha}(y;\boldsymbol{\theta},t_i,t_{i-1})\boldsymbol{u}(y;\boldsymbol{\theta},t_i,t_{i-1})dy\bigg]=\boldsymbol{0},
\end{equation}
where $\nabla$ represents gradient with respect to $\boldsymbol{\theta}$ and 
$\boldsymbol{u}(y;\boldsymbol{\theta},t_i,t_{i-1})$= $\nabla \ln f(y;\boldsymbol{\theta},t_i,t_{i-1})$ 
is the score function for the model of $i$-th increment density. 
Clearly, the estimating equation (\ref{Independent Increment Estimating Equation}) is unbiased 
when $g(\cdot;t_i,t_{i-1})$ belongs to the corresponding model family $\mathcal{F}_{i,\boldsymbol{t}}$ for all $i=1, \ldots, n$.

Observe that, in the limit as $\alpha \to 0$ the objective function  (plus $\frac{1}{\alpha}$) coverges to $1-\frac{1}{n}\sum_{i=1}^n\ln f(Y_i;\boldsymbol{\theta},t_i,t_{i-1})$,
which leads to the classical MLE having the estimating equation 
$\sum_{i=1}^n \boldsymbol{u}(Y_i;\boldsymbol{\theta},t_i,t_{i-1})=\boldsymbol{0}$.

\bigskip
\noindent\textbf{An Example (Drifted Brownian Motion):}\\
Recall that a wiener process $W(t)$ is defined as a continuous time stochastic process $\{W(t)\}_{t\geq 0}$ 
which satisfies the following properties: (i) $W(0)=0$, (ii) $W(t)$ has independent increments,
(iii) $W(t)$ is continuous in $t$, and (iv) $W(\cdot)$ has Gaussian increments, 
i.e., $W(t+u)-W(t)$ $\sim \mathcal{N}(0,u)$ for all $t, u>0$.

Then, we can define a drifted Brownian motion $X(t)$ with a mean functional  $\mu(t;\boldsymbol\theta)$ 
and a scale function $\sigma(\boldsymbol\theta)$ as 
\begin{equation*}
X(t)=\mu(t;\boldsymbol\theta)+\sigma(\boldsymbol\theta)W(t).
\end{equation*}
It is quite easy exercise to check that $\{X(t)\}_{t\geq 0}$ then belongs to the IIP family 
with the increment distribution being
\begin{equation*}
X(t)-X(s) \sim \mathcal{N}\left(\mu(t;\boldsymbol\theta)-\mu(s;\boldsymbol\theta),\sigma^2(\boldsymbol\theta)(t-s)\right).
\end{equation*}
For this case, the objective function for obtaining the MDPDE of $\boldsymbol{\theta}$ can be simplified as 
$$
H_{n, \alpha}(\boldsymbol{\theta};\boldsymbol{t}) = \frac{1}{n}\sum_{i=1}^n\frac{1}{\left(\sqrt{2\pi}\sigma_i(\boldsymbol{\theta})\right)^\alpha}\left[\frac{1}{\sqrt{1+\alpha}}-\left(1+\frac{1}{\alpha}\right)\exp\left(-\frac{\alpha(Y_i-\mu_i(\boldsymbol{\theta}))^2}{2\sigma_i^2(\boldsymbol{\theta})}\right)\right],
$$
where we define $Y_i=X_i - X_{i-1}$ for all $i=1, \ldots, n$, and
$$\sigma_i^2(\boldsymbol{\theta})=\sigma^2(\boldsymbol{\theta})(t_i-t_{i-1}) \quad \text{ and }\quad \mu_i(\boldsymbol{\theta})=\mu(t_i;\boldsymbol{\theta})-\mu(t_{i-1};\boldsymbol{\theta}).$$
Then, the corresponding estimating equation can be summarized as
\begin{align}
    \frac{1}{\left(\sqrt{2\pi}\sigma_i(\boldsymbol{\theta})\right)^\alpha}\times\bigg[&\sum_{i=1}^n \frac{\boldsymbol{\sigma'_i}(\boldsymbol{\theta})}{\sigma_i(\boldsymbol{\theta})}\left[\frac{1}{\sqrt{1+\alpha}}-\exp\left(-\frac{\alpha(Y_i-\mu_i(\boldsymbol{\theta}))^2}{2\sigma_i^2(\boldsymbol{\theta})}\right)\right]
    \nonumber\\
    &+\sum_{i=1}^n \left[\exp\left(-\frac{\alpha(Y_i-\mu_i(\boldsymbol{\theta}))^2}{2\sigma_i^2(\boldsymbol{\theta})}\right)\times \frac{(y-\mu_i(\boldsymbol{\theta}))\boldsymbol{\mu_i'}(\boldsymbol{\theta})}{\sigma_i(\boldsymbol{\theta})^2}\right]
    \nonumber\\
    &+\sum_{i=1}^n \frac{\boldsymbol{\sigma'_i}(\boldsymbol{\theta})}{\sigma_i(\boldsymbol{\theta})}\left[\exp\left(-\frac{\alpha(Y_i-\mu_i(\boldsymbol{\theta}))^2}{2\sigma_i^2(\boldsymbol{\theta})}\right)\frac{(Y_i-\mu_i(\boldsymbol{\theta}))^2}{\sigma_i^2(\boldsymbol{\theta})}-\frac{1}{(1+\alpha)^{3/2}\sigma_i(\theta)}\right]\bigg]=\boldsymbol{0},
    \label{EQ:example}
\end{align}
where $\boldsymbol{\mu_i'}(\boldsymbol{\theta})$ and $\boldsymbol{\sigma_i'}(\boldsymbol{\theta})$ are derivatives, with respect to $\boldsymbol{\theta}$,  
of $\mu_i(\boldsymbol{\theta})$ and $\sigma_i(\boldsymbol{\theta})$, respectively.
\hfill{$\square$}

\subsection{Extending the MDPDE for the Markov process}\label{Markov Process Estiamtor}
\label{SEC:MDPDE_MP}

Let us now consider a MP $\{X(t)\}_{t\geq 0}$ and the discretely observed sample $X_0, X_1, \ldots, X_n$ 
being the realizations of $\{X(t_i)\}_{0\leq i \leq n}$ 
for some $n\geq 1$ and pre-fixed time-stamp vector $\boldsymbol{t}$. 
For simplicity of notations, let us denote the conditional transition densities of the underlying MP 
as $g_{X_i|(X_{i-1}= y)}(x)=: g(x|y;t_i,t_{i-1})$.
Suppose that we want to model these conditional densities $g(\cdot|y;t_i,t_{i-1})$ by the family 
$\mathcal{F}_{i,y,\boldsymbol{t}}$ = $\{f(\cdot|y;\boldsymbol{\theta},t_i,t_{i-1});\boldsymbol{\theta} \in \Theta \}$ 
for all $i=1, 2, \ldots, n$ and for all $y$. 
Again observe that $f(\cdot|y;\boldsymbol{\theta},t_i,t_{i-1})$ can be very different across $i$ and $y$ 
but they share the same (vector) parameter $\boldsymbol{\theta}$ for every $y$ and $i$. 
However, since the observed data $\{X_i\}_{1\leq i\leq n}$ are dependent, here we propose to define the oracle MDPDE (associated functional), 
by extending the idea from \cite{ghosh2013robust}, as the minimizer of  the average expected DPD between 
the conditional densities of $X_i|(X_{i-1}=y)$  and its parametric model $ \mathcal{F}_{i,y,\boldsymbol{t}}$, which is 
\[
\frac{1}{n}\sum_{i=1}^n E_{X_{i-1}}(d_\alpha(g(X_i|X_{i-1};t_i,t_{i-1}),f(X_i|X_{i-1};\boldsymbol{\theta},t_i,t_{i-1}))).
\]
Given a set of observations, however, the MDPDE needs to be computed by minimizing an estimate of the above average discrepancy measure 
as given by 
$\frac{1}{n}\sum_{i=1}^n d_\alpha(\widehat{g}_i,f_i(\cdot|X_{i-1};\boldsymbol{\theta} ,t_i,t_{i-1}))$,
where $\widehat{g}_i$ is the density corresponding to the distribution degenerate at $X_i$. 
Since the third term of the divergence is independent of $\boldsymbol{\theta}$, again the MDPDE of $\boldsymbol{\theta}$ can be obtained by minimizing 
a simpler objective function given by 
\begin{equation}\label{Markov Objective Function}
\begin{split}
H_{n, \alpha}(\boldsymbol{\theta};\boldsymbol{t})=&\frac{1}{n}\sum_{i=1}^n\Bigg[\int f^{1+\alpha}(y|X_{i-1};\boldsymbol{\theta},t_i,t_{i-1})dy
-\left(1+\frac{1}{\alpha}\right)f^{\alpha}(X_i|X_{i-1};\boldsymbol{\theta},t_i,t_{i-1})\Bigg]\\
=&\frac{1}{n}\sum_{i=1}^n V_\alpha^M(X_i, X_{i-1}|\boldsymbol{\theta},t_i, t_{i-1}),
\end{split}
\end{equation}
where $V_\alpha^M(X_i, X_{i-1}|\boldsymbol{\theta},t_i, t_{i-1})$ indicates the term within square brackets
(superscript $M$ indicates that the underlying process has Markov property). 
Differentiating the above objective function we get the estimating equation for the DPD estimator in this setup as given by 
\begin{equation}\label{Markov Estimating Equation}
\begin{split}
\sum_{i=1}^n\bigg[f^{\alpha}(X_i|X_{i-1};\boldsymbol{\theta},t_i,t_{i-1})& \boldsymbol{u}(X_i| X_{i-1}; \boldsymbol{\theta},t_i,t_{i-1})-
\\  &\int f^{1+\alpha}(y|X_{i-1};\boldsymbol{\theta},t_i,t_{i-1})\boldsymbol{u}(y| X_{i-1},\boldsymbol{\theta},t_i,t_{i-1})dy\bigg]=\boldsymbol{0},
\end{split}
\end{equation}
where $\boldsymbol{u}(y| X_{i-1}; \boldsymbol{\theta},t_i,t_{i-1})$ = $\nabla \ln f(y|X_{i-1};\boldsymbol{\theta},t_i,t_{i-1})$. 
Note that the above estimating equation is again unbiased when the $i$-th conditional density $g(\cdot|X_{i-1};t_i,t_{i-1})$ 
belongs to the corresponding model family $\mathcal{F}_{i,X_{i-1},\boldsymbol{t}}$ for each realized value $X_{i-1}$
and all $i=1, \ldots, n$.

Note that, in the limit as $\alpha \to 0$, the minimizer of the MDPDE objective function also maximizes $\prod_{i=1}^n f(X_i|X_{i-1};\boldsymbol{\theta},t_i,t_{i-1})$ with respect to $\boldsymbol\theta$. 
Thus, the MDPDE with $\alpha \rightarrow 0$ is basically nothing but the MLE for which the estimating equation simplifies to
$\sum_{i=1}^n \boldsymbol{u}(X_i|X_{i-1}\boldsymbol{\theta},t_i,t_{i-1})=\boldsymbol{0}$.

\bigskip
\noindent\textbf{An Example (AR($1$) process with white noise errors):}\\
An AR($1$) process $X(t)$ is defined as a continuous time stochastic process $\{X(t)\}_{t\geq 0}$ satisfying
$$
X(t_i)=\mu +\rho X(t_{i-1}) +\epsilon(t_i),~~~ \mbox{with, }~ \epsilon(t_i)\sim \mathcal{N}(0,\sigma^2) ~~\mbox{for all } i=1, 2, \ldots,
$$
where $\mu\in \mathbb{R}$ and $\abs{\rho}\leq 1$ (to ensure the process is stationary). 
It is an easy exercise to check that $\{X(t_i)\}_{t\geq 0}$ then belongs to the MP family with the conditional distribution being
\begin{equation*}
	X(t_i)| X(t_{i-1}) \sim \mathcal{N}\left(\mu + \rho X(t_{i-1}),\sigma^2\right).
\end{equation*}
For this case, the objective function for obtaining the MDPDE of $\boldsymbol{\theta}=(\mu, \rho,\sigma^2)$ can be simplified as 
$$
H_{n, \alpha}(\boldsymbol{\theta};\boldsymbol{t}) = \frac{1}{n}\sum_{i=1}^n\frac{1}{\left(\sqrt{2\pi}\sigma\right)^\alpha}\left[\frac{1}{\sqrt{1+\alpha}}-\left(1+\frac{1}{\alpha}\right)
\exp\left(-\frac{\alpha(X(t_i)-\mu -\rho X(t_{i-1}))^2}{2\sigma^2}\right)\right].
$$
Then, the corresponding estimating equations are again given by (\ref{EQ:example})  with $Y_i = X(t_i)$, $\sigma_i^2(\boldsymbol{\theta})=\sigma^2$ and $\mu_i(\boldsymbol{\theta})=\mu + \theta X(t_{i-1}),$
so that 
$$
\boldsymbol{\mu_i'}(\boldsymbol{\theta}) = (1, X(t_{i-1}), 0), 
~~~~\mbox{ and }  ~~
\boldsymbol{\sigma_i'}(\boldsymbol{\theta}) =(0, 0, 1).
$$
\hfill{$\square$}

\biblio 

\section{Asymptotic Results}
\label{SEC:Asympt_results}

\subsection{A General Formulation: Main Theorem and Assumptions}

Let us combine the two definitions of MDPDE from Section 2 into a single setup. 
Recall the objective functions for both setups (IIP and MP models) have the same form but 
IIP involves the increment density $f(y;\boldsymbol{\theta},t_i,t_{i-1})$ and 
MP involves the transition densities $f(y|X_{i-1};\boldsymbol{\theta},t_i,t_{i-1})$,
while our parametric models are denoted by  $\mathcal{F}_{i,\boldsymbol{t}}$ and $F_{i,y,\boldsymbol{t}}$, respectively. 
In both cases, the MDPDE $\boldsymbol{\widehat{\theta}_n}$, at a given tuning parameter $\alpha\geq 0$,  
is defined by the relation (assuming that the minimum exists)
\begin{equation}\label{definition of thetag}
   H_{n, \alpha}(\boldsymbol{\widehat{\theta}_n};\boldsymbol{t})
   = \min_{\boldsymbol{\theta} \in \boldsymbol{\Theta}} H_{n, \alpha}(\boldsymbol{\theta};\boldsymbol{t}),
  ~~~~H_{n, \alpha}(\boldsymbol{\theta};\boldsymbol{t}) = \frac{1}{n}\sum_{i=1}^n V_i(\boldsymbol{\theta}),
\end{equation}
where only the form of the (stochastic) function $V_i(\boldsymbol{\theta})$ differs; 
for IIP models we have $V_i(\boldsymbol{\theta})=V_{\alpha}^I(Y_i|\boldsymbol{\theta},t_i, t_{i-1})$, 
whereas $V_i(\boldsymbol{\theta})=V_\alpha^M(X_i, X_{i-1}|\boldsymbol{\theta},t_i, t_{i-1})$ for the MP models.
So, we can consider (\ref{definition of thetag}) to be a general definition of the MDPDE for stochastic process models. 

Accordingly, if we denote $ H_\alpha^{(i)}(\boldsymbol{\theta})= E \left[V_i(\boldsymbol{\theta})\right]$ for $i=1, \ldots, n$,
then the population equivalent (functional) of the MDPDE is defined as 
\begin{equation}\label{stat functional}
\boldsymbol{\theta}_g=\min_{\boldsymbol\theta \in \Theta} \frac{1}{n} \sum_{i=1}^n H_\alpha^{(i)}(\boldsymbol{\theta}).
\end{equation}
Here, the expectation is taken with respect to the density of increment random variable $Y_i=X_i-X_{i-1}$ for the IIP model
and with respect to the joint density of $(X_i,X_{i-1})$ for the MP model;
the same notation of expectation (and also for variance) will be assumed 
throughout the rest of the paper unless mentioned otherwise. 
Further, throughout our theoretical discussions, we will assume that $\boldsymbol{\theta}_g$ exists
and refer it to as the best fitting parameter value (in the DPD sense).
This assumption clearly holds if the true density belongs to the model family for some common vector parameter value
$\boldsymbol{\theta}=\boldsymbol{\theta}_0$ where the best fitting parameter $\boldsymbol{\theta}_g$ 
becomes the true parameter value $\boldsymbol{\theta}_0$.
Note that, at the best fitting parameter $\boldsymbol{\theta}_g$ defined by (\ref{stat functional}), we have
$    \sum_{i=1}^n \nabla H_\alpha^{(i)}(\boldsymbol{\theta}_g)= \boldsymbol{0}_p$.
We assume that the individual terms in the above sum is also zero for $i=1, \ldots, n$,
which implies that the same $\boldsymbol{\theta}_g$ minimizes the individual $H_\alpha^{(i)}(\boldsymbol{\theta})$ for all $i$.

Next, we define two $p \times p$ matrices, for the variance computation, as 
\begin{equation}\label{Psi and Omega}
\boldsymbol{\Omega}_{n}(\boldsymbol{t})= \frac{1}{n}Var\left[\sum_{i=1}^n \boldsymbol{\nabla} V_i(\boldsymbol{\theta})\right],
 \quad \text{ and } \quad 
 \boldsymbol{\Psi}_{n}(\boldsymbol{t})= \frac{1}{n}\sum_{i=1}^n \boldsymbol{J}^{(i)}(\boldsymbol{t}),
\end{equation}
with the matrix $\boldsymbol{J}^{(i)}(\boldsymbol{t}$) having the $(k,l)$-th entry given by
$\boldsymbol{J}_{kl}^{(i)}(\boldsymbol{t})= E \left[\nabla_{kl}(V_i(\boldsymbol{\theta}))\right]$ for all $k,l =1, \ldots, p$,
where $\boldsymbol{\nabla}$ denotes the (first order) gradient with respect to $\boldsymbol{\theta}$, $\nabla_{kl}$ represents 
the second order partial derivative with respect to the indicated components ($k, l$) of $\boldsymbol{\theta}$ 
and the associated expectation and variance are considered with respect to different densities for the IIP and MP set-ups 
as described above. 
In particular, for the IIP models, the variance can be taken inside the summation in the expression of $\boldsymbol\Omega_n(\boldsymbol t)$ in (\ref{Psi and Omega}).

Now our goal is to establish the asymptotic normality and consistency of the MDPDE for both the IIP and MP models. 
Our main theorem in this respect, under the above-mentioned general set-up, is described below.

\begin{theorem}[Main Result]\label{main theorem}
Consider the discretely observed data at pre-fixed time stamp vector $\boldsymbol{t}$ from a stochastic process model,
which is either IIP or MP. Let $\alpha\geq 0$ is fixed. Then under suitable assumptions (stated below), 
we will have the following results. 
\begin{enumerate}
\item There exists a consistent sequence $\boldsymbol{\widehat{\theta}}_n$ of roots of the minimum DPD estimating equations.
\item The asymptotic distribution of 
$\boldsymbol{\Omega}_n(\boldsymbol{t})^{-\frac{1}{2}}\boldsymbol{\Psi}_n(\boldsymbol{t})
[\sqrt{n}(\boldsymbol{\widehat{\theta}}_n -\boldsymbol{\theta}_g)]$ is p-dimensional normal 
with (vector) mean $\boldsymbol{0}_p$ and co-variance matrix $\boldsymbol{I}_p$, the $p$-dimensional identity matrix.
\end{enumerate}
\end{theorem}


In consistent with the general notation considered so far in this section, 
the common set of assumptions required to prove Theorem \ref{main theorem} are listed below,
where $f_i(y)$ denotes $f(y;\boldsymbol{\theta},t_i,t_{i-1})$ or $f(y|X_{i-1};\boldsymbol{\theta},t_i,t_{i-1})$ 
for the IIP or the MP models, respectively, and $g_i(y)$ denotes the corresponding population density in each case.   
 
\bigskip\noindent
\textbf{Common Set of Assumptions:} 
\setlist[condenum]{label=(\bfseries A\arabic*), ref=A\arabic*, wide}
\begin{condenum}
	\item The support $\mathcal{X}=\{y|f_i(y)>0 \}$ is independent of $i$ and $\boldsymbol{\theta}$ for all $i$, 
	the true densities $g(\cdot;t_i,t_{i-1})$ are also supported on $\mathcal{X}$ for all $i$.	
	\label{Indep ass 1}

	\item There is an open subset of the parameter space $\boldsymbol{\Theta}$, 
	containing the best fitting parameter $\boldsymbol{\theta}_g$ such that for almost all $y\in \mathcal{X}$, $\boldsymbol{\theta} \in \boldsymbol{\Theta}$, and all $i=1,2,\ldots$,
	the density $f_i(y)$ is thrice differentiable with respect to $\boldsymbol{\theta}$ 
	and the third partial derivatives are continuous with respect to $\boldsymbol{\theta}$.\label{Indep ass 2}
	
	\item For $i=1,2,\ldots$ the integrals $\int f_i^{1+\alpha}(y;\boldsymbol\theta)dy$ and $\int f_i^\alpha(y;\boldsymbol\theta)g_i(y)dy$ 
	can be differentiated thrice with respect to $\boldsymbol{\theta}$, 
	and the derivatives can be taken under the integral sign.\label{Indep ass 3}
	
	\item For each $i=1,2,\ldots$ the matrices $\boldsymbol{J}^{(i)}(\boldsymbol{t})$ are positive definite and
	\begin{equation*}
	\lambda_0(\boldsymbol{t})=\inf_{n} \text{[min eigenvalue of $\boldsymbol{\Psi}_n(\boldsymbol{t})$]}>0.
	\end{equation*}\label{Indep ass 4}
\end{condenum}

Note that, Assumptions (\ref{Indep ass 1})--(\ref{Indep ass 4}) are quite standard in the theory of asymptotic inference
and are satisfied by most common regular statistical model distributions. 
However, we also need a few more assumptions towards achieving the asymptotic properties of the MDPDE,
as mentioned in Theorem \ref{main theorem}, which are specific to the model set-ups as described before. 
These assumptions involve the boundedness of the third derivatives of the respective objective functions,
requirements for the laws of large numbers in respective set-ups and 
also the Lindberg-Feller type conditions to apply the corresponding central limit theorem (CLT)
for proving the asymptotic normality results. 
They will be further simplified for several important special cases of IIPs and MPs in the subsequent subsections. 
At this moment, we also define $I(\cdot)$ to represent Indicator function.

\bigskip\noindent
\textbf{Specific Assumptions required for the IIP Model:}
\setlist[condenum]{label=(\bfseries I\arabic*), ref=I\arabic*, wide}
\begin{condenum}
    \item There exist functions $M^{(i)}_{jkl}(y)$ such that, for all $i=1,2, \ldots$ and all $j,k,l=1, \ldots, p$,
\begin{align*}
&\left|\nabla_{jkl}V_{\alpha}^I(y|\boldsymbol{\theta},t_i, t_{i-1})\right| \leq M^{(i)}_{jkl}(y), 
\quad \text{ for all } \boldsymbol{\theta} \in \boldsymbol{\Theta}, y\in\mathcal{X},  \text{ with }~~
\frac{1}{n}\sum_{i=1}^n E\bigg[M^{(i)}_{jkl}(Y_i)\bigg]= O(1).
        \end{align*}\label{Indep ass 5}

    \item For all $j, k = 1, \ldots, p$, we have
    \begin{equation*}
\lim_{N\to \infty} \sup_{n>1} \bigg\{\frac{1}{n} \sum_{i=1}^n 
E\left[|\nabla_j V_{\alpha}^I(Y_i|\boldsymbol{\theta},t_i, t_{i-1})|
I\left(|\nabla_j V_{\alpha}^I(Y_i|\boldsymbol{\theta},t_i, t_{i-1})|>N\right)\right]\bigg\}= 0,
    \end{equation*}
   and
    \begin{align*}
\lim_{N\to \infty} \sup_{n>1} \bigg\{\frac{1}{n} \sum_{i=1}^n 
& E\left[|\nabla_{jk} V_{\alpha}^I(Y_i|\boldsymbol{\theta},t_i, t_{i-1}) 
-E(\nabla_{jk}V_{\alpha}^I(Y_i|\boldsymbol{\theta},t_i, t_{i-1}))|\right.\\
&\left. I\left(|\nabla_{jk} V_{\alpha}^I(Y_i|\boldsymbol{\theta},t_i, t_{i-1})
-E(\nabla_{jk}V_{\alpha}^I(Y_i|\boldsymbol{\theta},t_i, t_{i-1}))|>N\right)\right] \bigg\}= 0.
    \end{align*} \label{Indep ass 6}

    \item For all $\epsilon >0$, we have
    \begin{align*}
\lim_{n\to \infty} \bigg\{\frac{1}{n} \sum_{i=1}^n 
E\left[\|\boldsymbol{\Omega}_n(\boldsymbol{t})^{-1/2}\nabla V_{\alpha}^I(Y_i|\boldsymbol{\theta},t_i, t_{i-1})\|^2
I\left(\|\boldsymbol{\Omega}_n(\boldsymbol{t})^{-1/2}
\nabla V_{\alpha}^I(Y_i|\boldsymbol{\theta},t_i, t_{i-1})\|>\epsilon\sqrt{n}\right)\right]\bigg\}= 0.
    \end{align*} \label{Indep Ass 7}
\end{condenum}

\bigskip\noindent
\textbf{Specific Assumptions required for the MP Model:}
\setlist[condenum]{label=(M\arabic*), ref=M\arabic*, wide}
\begin{condenum}

\item There exist functions $M^{(i)}_{jkl}(y,z)$ such that, for all $i=1,2, \ldots$ and all $j,k,l=1, \ldots, p$,
\begin{align*}
&\left|\nabla_{jkl}V_\alpha^M(y,z|\boldsymbol{\theta},t_i, t_{i-1})\right| \leq M^{(i)}_{jkl}(y,z), 
\quad \text{ for all } \boldsymbol{\theta} \in \boldsymbol{\Theta}, y,z\in\mathcal{X},  \text{ with }~~
\frac{1}{n}\sum_{i=1}^n E\bigg[M^{(i)}_{jkl}(X_i, X_{i-1})\bigg]= O(1).
\end{align*}
Further, for all $i,j,k,l$, $E[M_{jkl}^{(i)}(X_i,X_{i-1})]^2$ exists finitely
and there exist $\sigma^2_{jkl}\in(0,\infty)$ and $c_{jkl}\in[0,1]$ such that
\[
Cov\left(M_{jkl}^{(i_1)}(X_{i_1},X_{i_1-1}),M_{jkl}^{(i_2)}(X_{i_2},X_{i_2-1})\right)\leq \sigma_{jkl}^2c_{jkl}^{|i_1-i_2|},
~~~\mbox{for all } i_1, i_2 = 1, 2, \ldots.
\]
\label{MP ass 1}

\item For every $k=1, \ldots, p$, $E[\nabla_k V_\alpha^M(X_i, X_{i-1}|\boldsymbol{\theta},t_i, t_{i-1})]^2$ exists finitely
for every $i=1, 2, \ldots$, and there exists constants $0\leq c_{1,k}\leq 1$ and $\sigma_{1,k}>0$ such that
\[
Cov\left(\nabla_k V_\alpha^M(X_{i_1},X_{i_1-1}|\boldsymbol{\theta},t_i, t_{i-1}),
\nabla_k V_\alpha^M(X_{i_2},X_{i_2-1}|\boldsymbol{\theta},t_i, t_{i-1})\right)\leq \sigma_{1,k}^2c_{1,k}^{|i_1-i_2|},
~~~\mbox{for all } i_1, i_2 = 1, 2, \ldots.
\]
Further, for every $j, k = 1, \ldots, p$, 
$E[\nabla_{jk} V_\alpha^M(X_i, X_{i-1}|\boldsymbol{\theta},t_i, t_{i-1})]^2$ exists finitely for every $i=1,2, \ldots$,
and there exists constants $0\leq c_{1,j,k}\leq 1$ and $\sigma_{1,j,k}>0$ such that
\[
Cov\left(\nabla_{jk} V_\alpha^M(X_{i_1},X_{i_1-1}|\boldsymbol{\theta},t_i, t_{i-1}),
\nabla_{jk} V_\alpha^M(X_{i_2},X_{i_2-1}|\boldsymbol{\theta},t_i, t_{i-1})\right)\leq \sigma_{1,j,k}^2c_{1,j,k}^{|i_1-i_2|},
~~~\mbox{for all } i_1, i_2 = 1, 2, \ldots.
\]
\label{MP ass 2}

\item There exists a positive definite matrix $\boldsymbol\Omega(\boldsymbol{t})$ such that 
$\boldsymbol\Omega_n(\boldsymbol{t})$ converges element-wise to $\boldsymbol\Omega(\boldsymbol{t})$ 
as $n\to \infty$. 
Further, the sequence $\left\{ V_\alpha^M(X_i, X_{i-1}|\boldsymbol{\theta},t_i, t_{i-1}) \right\}_{i\geq 1}$ 
is strictly stationary and satisfies strong mixing condition (as defined below in Definition \ref{Def:mixing_strong})
with $\alpha_n=O(n^{-5}))$, and 
\[
E\left[\nabla_j V_\alpha^M(X_i, X_{i-1}|\boldsymbol{\theta},t_i, t_{i-1})\right]^{12} \leq \infty,
~~~~\mbox{ for all } j =1, \ldots, p \text{ and } i=1,2,\cdots
\]
\label{MP ass 3}
\end{condenum}

\begin{definition}[Strong Mixing Condition]
For a sequence $Z_1, Z_2, \ldots$ of random variables, let $\alpha_n$ denote a number such that
\[|P(A\cap B)-P(A)P(B)|\leq \alpha_n \]
for every $A\in \sigma(Z_1, Z_2,\ldots, Z_k)$, $B\in \sigma(Z_{k+n}, Z_{k+n+1}, \ldots)$ with $k\geq 1$ and $n\geq 1$. 
The sequence is said to satisfy strong mixing condition if $\alpha_n\to 0$ as $n\to \infty$. 
Intuitively speaking, under strong mixing condition, $Z_k$ and $Z_{k+n}$ are approximately independent for large enough $n$ and any $k$.
\label{Def:mixing_strong}
\end{definition}

\bigskip
Now, we can prove Theorem \ref{main theorem} for the IIP and MP model set-ups under the aforementioned assumption.
For the IIP set-ups, under Assumptions (\ref{Indep ass 1})--(\ref{Indep ass 4}) and (\ref{Indep ass 5})--(\ref{Indep Ass 7}),
the proof of the consistency and asymptotic normality of the MDPDEs (as stated in Theorem \ref{main theorem}) 
can be done relatively easily by following the general theory under non-homogeneous observations from \cite{ghosh2013robust};
we leave it as an exercise for the readers. 
However, the proof of Theorem \ref{main theorem} under the MP set-ups requires non-trivial extensions of the existing arguments,
which is presented in the following subsection. 

\subsection{Proof of Theorem \ref{main theorem} for the MP models}

Before going to the proof of Theorem \ref{main theorem} regarding the asymptotic properties of the MDPDE under the MP set-ups,
let us state couple of important lemmas which will be instrumental in the proof of our main theorem.
The proof of the first lemma is given in Appendix \ref{lemma proofs}, 
whereas we refer the readers to \cite{billingsley2008probability} for a proof of the second lemma. 

\bigskip
\begin{lemma}\label{markov process wlln lemma}
Consider a set of dependent  observations $\{Z_i\}_{i\geq 0}$ such that $E(Z_i^2)$ exists finitely for all $i$. 
Let $S_n$ be the partial sum up to $Z_n$. Suppose that there exists $\sigma^2>0$ and $0\leq c\leq 1$ such that
\[
|Cov(Z_i, Z_j)|\leq \sigma^2 c^{|i-j|} \quad 
\mbox{for all }  i, j=1, 2, \ldots.
\]
Then, it follows that 
$
n^{-1}\left[{S_n-E(S_n)}\right]\overset{P}{\to}0
$
as $n\to \infty$. 
\end{lemma}

\begin{lemma}[\cite{billingsley2008probability}]\label{markov process clt}
Suppose that $\{Z_i\}_{i\geq 0}$  is a strictly stationary sequence of random variables
which satisfies the strong mixing condition with $\alpha_n=O(n^{-5})$ and 
that $E(Z_n)=0$ and $EZ_n^{12}\leq \infty$ for every $n\geq 1$. 
If $S_n$ denotes the partial sum up to $Z_n$, then
\[
n^{-1}Var(S_n)\to \sigma^2:=E(Z_1^2)+2\sum_{k=1}^\infty E(Z_1Z_{k+1}),
\]
as $n\to \infty$, where the series (in the definition of $\sigma^2$) indeed converges absolutely. 
Further, if $\sigma>0$, then 
\[
\frac{S_n}{\sigma\sqrt{n} }\overset{d}{\to} \mathcal{N}(0,1),
~~~~\mbox{as }~n \to \infty.
\]
\end{lemma}

\bigskip\noindent
\textbf{Proof of Part 1 of Theorem \ref{main theorem} (Consistency):}\\
We will follow the argument for proving the consistency of the MDPDE under IID set-ups from \cite{Basu/etc:2011},
along with the necessary modifications required for the present (dependent) MP set-ups. 
To prove the consistency, we will consider the behaviour of the density power divergence or equivalently $H_{n, \alpha}(\boldsymbol{\theta}):=H_{n, \alpha}(\boldsymbol{\theta};\boldsymbol{t})$ on a sphere $Q_a$ with the center at the best fitting parameter $\boldsymbol{\theta}^g$ and radius $a$  (Here, we are suppressing $\boldsymbol{t}$ in the notation for simplicity).

We will show that for any sufficiently small $a$ the probability tends to 1 that $H_{n, \alpha}(\boldsymbol\theta)>H_{n, \alpha}(\boldsymbol\theta_g)$ for all points $\boldsymbol\theta$ on the surface $Q_a$ and hence that $H_{n, \alpha}(\boldsymbol\theta)$ has a local minimum in the interior of $Q_a$. Since at a local minimum the estimating equation must be satisfied, it will follow that for any $a>0$ with probability tending to $1$ as $n\to \infty$ the estimating equation (\ref{Markov Estimating Equation}) has a solution $\widehat{\boldsymbol\theta}_n(a)$ within $Q_a$. 

To study the behaviour of $H_{n, \alpha}(\boldsymbol\theta)$ on $Q_a$,we expand $H_{n, \alpha}(\boldsymbol\theta)$ by the Taylor series as follows
\begin{align*}
H_{n, \alpha} (\boldsymbol\theta^g)- H_{n, \alpha}(\boldsymbol\theta)
=&\bigg[\underbrace{\sum_j (-A_j)(\theta_j-\theta_j^g)}_{S_1}+\underbrace{\frac{1}{2}\sum_j\sum_k (-B_{jk})(\theta_j-\theta_j^g)(\theta_k-\theta_k^g)}_{S_2}\\
+&\underbrace{\frac{1}{6}\sum_j\sum_k\sum_l (\theta_j-\theta_j^g)(\theta_k-\theta_k^g)(\theta_l-\theta_l^g)\frac{1}{n}\sum_{i=1}^n \gamma_{jkl}(X_i,X_{i-1})M_{jkl}^{(i)}(X_i,X_{i-1})}_{S_3}\bigg]
\end{align*}
where, we have
\begin{align*}
A_j=&\boldsymbol\nabla_j H_{n, \alpha}(\boldsymbol\theta)|_{\boldsymbol\theta=\boldsymbol\theta^g}=\frac{1}{n}\sum_{i=1}^n \boldsymbol\nabla_j V_\alpha^M(X_i, X_{i-1}|\boldsymbol{\theta},t_i, t_{i-1})\bigg|_{\boldsymbol\theta=\boldsymbol\theta^g},\\
B_{jk}=&\boldsymbol\nabla_{jk} H_{n, \alpha}(\boldsymbol\theta)|_{\boldsymbol\theta=\boldsymbol\theta^g}
=\frac{1}{n}\sum_{i=1}^n \boldsymbol\nabla_{jk} V_\alpha^M(X_i, X_{i-1}|\boldsymbol{\theta},t_i, t_{i-1})\bigg|_{\boldsymbol\theta=\boldsymbol\theta^g},
\end{align*}
with $\boldsymbol\nabla_j$ and $\boldsymbol\nabla_{jk}$ representing the partial derivatives with the indicated components of $\boldsymbol\theta$ and $0\leq |\gamma_{jkl}(x)|\leq 1$.

Now we start with noting that, for each $i,j$,
$$
E_{X_i,X_{i-1}}\left(\boldsymbol\nabla_j V_\alpha^M(X_i, X_{i-1}|\boldsymbol{\theta},t_i, t_{i-1})\right)\bigg|_{\boldsymbol\theta=\boldsymbol\theta^g}=\boldsymbol\nabla_j H^{(i)}(\boldsymbol\theta_g)=0.
$$
Now, for all $j$, using Condition $1$ of (\ref{MP ass 2}) and Lemma \ref{markov process wlln lemma}, we have
$$A_j\overset{P}{\to}0.$$
Then, for any given $a$, we have $|A_j|< a^2 \text{ a.e.}$, and hence, $|S_1|<pa^3$ with probability tending to $1$ as $n\to\infty$.

Similarly, observing the definition of $\Psi_n$ one can show, for all $j,k$ using Lemma \ref{markov process wlln lemma} and Condition $2$ of (\ref{MP ass 2}), that
\[
 B_{jk}-(\boldsymbol\Psi_n)_{jk}\overset{P}{\to}0.
\]
Now, consider the representation
\begin{align*}
2S_2=\sum_j & \sum_k \left[-(\boldsymbol\Psi_n)_{jk}(\theta_j-\theta^g_j)(\theta_k-\theta_k^g)\right]
+ \sum_j\sum_k \left[\left(- B_{jk}+(\boldsymbol\Psi_n)_{jk}\right)(\theta_j-\theta^g_j)(\theta_k-\theta_k^g)\right]
\end{align*}
For the second term in the above equation it follows, form an argument similar to that for $S_1$, that it's absolute value is less than $p^2a^3$ with probability tending to $1$ as $n\to\infty$.

The first term is a negative (non-random) quadratic form in the variables $(\theta_j-\theta^g_j)$. 
By an orthogonal transformation this can be reduced to a diagonal form $\sum_i \lambda_i\xi_i^2$ with $\sum_i\xi_i^2=a^2$.
The quantities $\lambda_i$ and $\xi_i$ are also functions of $n$ which has been suppressed here for brevity. As each $\lambda_i$ is negative, by ordering them and using Assumption (\ref{Indep ass 4}), we get
$$\sum_i\lambda_i\xi_i^2\leq -\lambda_0 a^2$$.
Combining the first and the second terms, thus, there exist $c>0$ and $a_0>0$ such that $S_2<-ca^2 \text{ a.e.}$, for all $a<a_0$, as $n\to\infty$.

Finally, using Assumption (\ref{MP ass 1}), we have some $0<m_{jkl}<\infty$ for which
\[
\bigg|\frac{1}{n} \sum_i \gamma_{jkl}(X_i,X_{i-1}) M_{jkl}^{(i)}(X_i,X_{i-1})\bigg|\leq \bigg|\frac{1}{n} \sum_i M_{jkl}^{(i)}(X_i,X_{i-1})\bigg|<m_{jkl}<\infty \quad \text{a.e.},
\]
as $n\to\infty$, and hence, $|S_3|<ba^3$ on $Q_a$, where
$b=\frac{1}{6}\sum_j\sum_k\sum_l m_{jkl}.$
Combining all these, we get
\[
\max(S_1+S_2+S_3)<-ca^2+(b+s)a^3,
\]
which is less than $0$ if $a<\frac{c}{b+s}$. Then for sufficiently small $a$ there exists a sequence of roots $\widehat{\boldsymbol\theta}=\widehat{\boldsymbol\theta}(a)$ such that
\[
P(\|\widehat{\boldsymbol\theta}-\boldsymbol\theta^g\|<a)\to 1 \quad \text{as} \quad n\to\infty.
\]

It remains to show that we can determine such a sequence independently of $a$.
Let $\boldsymbol\theta^\star$ be the root closest to $\boldsymbol\theta^g$. 
This exists because the limit of a sequence of roots is again a root by the continuity of $H_{n, \alpha}(\boldsymbol\theta)$ as a function of $\boldsymbol\theta$.
Then, clearly
\[
P(\|\boldsymbol\theta^\star-\boldsymbol\theta^g\|<a)\to 1 \quad \forall a>0.
\]
This conclude the proof of the existence of a sequence of constant solution to the estimating equation with probability tending to $1$ as $n\to \infty$.

\bigskip\noindent
\textbf{Proof of Part 2 of Theorem \ref{main theorem} (Asymptotic Normality):}\\
Note that, in view of Assumption (M3), it is equivalent to proof the asymptotic normality of $\widehat{\boldsymbol\theta}_n$ with $\boldsymbol\Omega_n(\boldsymbol t)$ replaced by $\boldsymbol\Omega(\boldsymbol t)$.
For this purpose, let us expand $H_{n, \alpha}^j(\boldsymbol\theta):=\nabla_j H_{n, \alpha}(\boldsymbol\theta)$ about $\boldsymbol\theta^g$ as follows:

\begin{align*}
H_{n, \alpha}^j(\boldsymbol\theta)=H_{n, \alpha}^j(\boldsymbol\theta^g)&+\sum_k (\theta_k-\theta_k^g)H_{n, \alpha}^{jk}(\boldsymbol\theta^g)
+\frac{1}{2}\sum_k\sum_l (\theta_k-\theta_k^g)(\theta_l-\theta_l^g)H_{n, \alpha}^{jkl}(\boldsymbol\theta^\star),
\end{align*}
where $\boldsymbol\theta^\star$ is a point on the line segment joining $\boldsymbol\theta$ and $\boldsymbol\theta^g$ and
$H_{n, \alpha}^{jk}$ and $H_{n, \alpha}^{jkl}$ denote the corresponding second and third partial derivatives of $H_{n, \alpha}$. 
But since $H_{n, \alpha}^j(\widehat{\boldsymbol\theta})=0$, we get (here we are suppressing $n$ in the notation of $\widehat{\boldsymbol\theta}_n$)
\begin{align*}
n^{1/2}\sum_k (\widehat{\theta}_k-\theta_k^g)\left[H_{n, \alpha}^{jk}(\boldsymbol\theta^g)+\frac{1}{2}\sum_{l=1}^p(\widehat{\theta}_l-\theta_l^g)H_{n, \alpha}^{jkl}(\boldsymbol\theta^\star)\right]
=n^{1/2}H_{n, \alpha}^j(\boldsymbol\theta^g).
\end{align*}
Above equation can also be expressed as
\begin{align*}
\sum_k A_{jkn}Z_{kn}=T_{jn}, ~~\mbox{ for all } j=1, \ldots, p, ~~\mbox{ i.e., }~~ \boldsymbol A_n \boldsymbol Z_n=\boldsymbol T_n,
\end{align*}
where, we define $\boldsymbol T_n=(T_{1n},\hdots , T_{pn})$, $\boldsymbol Z_n=(Z_{1n},\hdots , Z_{pn})$ and $\boldsymbol A_n=((A_{jkn}))_{j=1,\hdots,p;k=1,\hdots,p}$, with
\begin{align*}
T_{jn}=& - n^{1/2}H_{n, \alpha}^j(\boldsymbol\theta^g), ~~~~~~~~~~~~~~Z_{kn}=n^{1/2} (\widehat{\theta}_k-\theta_k^g),\\
\mbox{ and } ~~ A_{jkn}=&H_{n, \alpha}^{jk}(\boldsymbol\theta^g)+\frac{1}{2}\sum_{l=1}^p(\widehat{\theta}_l-\theta_l^g)H_{n, \alpha}^{jkl}(\boldsymbol\theta^\star).
\end{align*}

Now, we note that $\boldsymbol T_n\in \mathbb{R}^p$ can be expressed as
\[
\boldsymbol T_n=\frac{1}{\sqrt{n}}\sum_{i=1}^n\nabla V_\alpha^M(X_i, X_{i-1}|\boldsymbol{\theta},t_i, t_{i-1}). 
\]
Also, observe that $E(\nabla V_\alpha^M(X_i, X_{i-1}|\boldsymbol{\theta},t_i, t_{i-1}))=\boldsymbol{0}$.
Take any non-zero vector $\boldsymbol a\in \mathbb{R}^p$ and consider
$$Y_i=\boldsymbol a^T\nabla V_\alpha^M(X_i, X_{i-1}|\boldsymbol{\theta},t_i, t_{i-1}).$$
Observe that, by Assumption (\ref{MP ass 3}), $Y_1,Y_2,\hdots$ is a strictly stationary sequence of random variables. Also, for any $k$, we have
\[
\sigma(Y_1,Y_2,\hdots,Y_k)\subseteq \sigma\left(\nabla V_\alpha^M(X_1, X_0|\boldsymbol{\theta},t_1, t_0),\hdots,\nabla V_\alpha^M(X_k, X_{k-1}|\boldsymbol{\theta},t_k, t_{k-1})\right).
\]
Thus, by definition \ref{Def:mixing_strong} and Assumption (\ref{MP ass 3}), we can verify that $Y_1,Y_2,\hdots$ also satisfy strong mixing conditions with $\alpha_n=O(n^{-5})$.

Now, let us define $S_n=Y_1+Y_2+\hdots+Y_n$ and note $S_n=\sqrt{n}\boldsymbol a^T\boldsymbol T_n$. It is easy to check that
\[
Var(S_n)=Var\left(\boldsymbol a^T\sum_{i=1}^n \nabla V_\alpha^M(X_i, X_{i-1}|\boldsymbol{\theta},t_i, t_{i-1})\right)=\boldsymbol a^T\boldsymbol\Omega_n(\boldsymbol{t})\boldsymbol a.
\]
But, $\boldsymbol a^T\boldsymbol \Omega_n(\boldsymbol{t})\boldsymbol a\to \boldsymbol a^T\Omega(\boldsymbol{t})\boldsymbol a > 0$ from Assumption (\ref{MP ass 3}). 
Hence, by theorem \ref{markov process clt}, we  have
\[
\frac{\boldsymbol a^T \boldsymbol T_n}{\sqrt{\boldsymbol a^T\boldsymbol \Omega(\boldsymbol{t})\boldsymbol a}}\overset{d}{\to}\mathcal{N}(0,1).
\]
Since $\boldsymbol a$ was arbitrary non-zero vector, we use Cramer-Wold theorem to conclude that
$\boldsymbol\Omega(\boldsymbol{t})^{-1/2}\boldsymbol T_n\overset{D}{\to}\mathcal{N}_p(0,\boldsymbol I_p)$ as $n\rightarrow\infty$, 
which can also be expressed as
\begin{eqnarray}
 \boldsymbol\Omega(\boldsymbol{t})^{-1/2}\boldsymbol A_n\boldsymbol Z_n\overset{D}{\to}\mathcal{N}_p(0,\boldsymbol I_p).
\label{EQ:star}
\end{eqnarray}

Now, as we have shown in the proof of consistency above, using Assumption (\ref{MP ass 1}), $H_{n, \alpha}^{jkl}(\boldsymbol\theta^\star)$ is bounded a.e. as $n\to\infty$, 
and hence, consistency of $\widehat{\boldsymbol\theta}$ implies that second term of $A_{jkn}$ converges to $0$ in probability as $n\to\infty$ for all $j,k$.
Proceeding as in proof of the consistency,  we have
$H_{n, \alpha}^{jk}(\boldsymbol\theta^g)-(\boldsymbol\Psi_n)_{jk} \overset{P}{\to}0$  for all $j, k$.
Thus, it follows that
\[
\boldsymbol\Omega(\boldsymbol{t})^{-1/2}\left[\boldsymbol A_n-\boldsymbol\Psi_n\right]\boldsymbol Z_n\overset{P}{\to}0.
\]

Finally, using Slutsky's theorem along with (\ref{EQ:star}), we have
\[
\boldsymbol\Omega(\boldsymbol{t})^{-1/2}\boldsymbol\Psi_n \boldsymbol Z_n \overset{D}{\to} \mathcal{N}_p(0,\boldsymbol I_p).
\]
\hfill{$\square$}

\subsection{Special Case I: Single Family of Distributions in IIP}
\label{SEC:single family}

We have seen that we need some non-trivial Assumptions like (\ref{Indep ass 6}) and (\ref{Indep Ass 7}) to achieve asymptotic results for MDPDE in the IIP setup. But we can relax or replace those assumptions for most practical IIP set-ups.
Let us start with the special case of the IIP set-ups 
where all the increment distributions come from a single family of distribution having density $f$ 
(with possibly different parameters, say 
$\boldsymbol\lambda_i=\boldsymbol\lambda_i(\boldsymbol\theta)\in \Lambda\subseteq\mathbb{R}^\ell$)
and we also assume that our model is correct i.e. all the true densities come from the respectively model family 
at some (true) parameter value $\boldsymbol{\theta}_0=\boldsymbol{\theta}_g$.
The common practical examples within this subclass include Poisson process and drifted Brownian motion, 
where the common family of increment distributions are Poisson and normal, respectively.

In terms of our general notations, we now have 
$f(y;\boldsymbol{\theta},t_i,t_{i-1})= f(y,\boldsymbol\lambda_i(\boldsymbol{\theta}))$,
and hence, 
\begin{equation*}
    V_{\alpha}^I(Y_i|\boldsymbol{\theta},t_i, t_{i-1})= \int f^{1+\alpha}(y,\boldsymbol\lambda_i)dy-\left(1+\frac{1}{\alpha}\right)f^\alpha(Y_i,\boldsymbol\lambda_i).
\end{equation*}
Let us define $\boldsymbol u_\lambda(y,\boldsymbol\lambda)=\boldsymbol \nabla_{\boldsymbol\lambda}\ln f(y,\boldsymbol\lambda)$ 
as the score function corresponding to the density $f$, 
where $\boldsymbol\nabla_{\boldsymbol\lambda}$ represents the (first order) gradient with respect to $\boldsymbol{\lambda}\in\Lambda$,
and then also define
\begin{equation*}
    \begin{split}
\boldsymbol C_{\alpha}(\boldsymbol\lambda) 
=\int f^{1+\alpha}(y,\boldsymbol\lambda)\boldsymbol u_\lambda(y,\boldsymbol\lambda)dy,&
~~~~ 
\boldsymbol C^{(1)}_{\alpha}(\boldsymbol\lambda)
=\int f^{1+\alpha}(y,\boldsymbol\lambda)\boldsymbol\nabla_{\boldsymbol\lambda}\boldsymbol u_\lambda(y,\boldsymbol\lambda)dy, 
\\ 
\mbox{and }~~ \boldsymbol C^{(2)}_{\alpha}(\boldsymbol\lambda)
=\int &f^{1+\alpha}(y,\boldsymbol\lambda)\boldsymbol u_\lambda(y,\boldsymbol\lambda)
\boldsymbol u_\lambda(y,\boldsymbol\lambda)^Tdy,
\end{split}
\end{equation*}
where the integrations are done element wise.
%
%
Then, after some simplification, for any $j, k=1, \ldots, p$, one can show the following results:
\begin{align*}
\boldsymbol\nabla_j V_{\alpha}^I(Y_i|\boldsymbol{\theta},t_i, t_{i-1}) 
&=(1+\alpha)\left(\pdv{\boldsymbol\lambda_i}{\theta_j}\right)^T[\boldsymbol C_{\alpha}(\boldsymbol\lambda_i)-f^\alpha(Y_i,\boldsymbol \lambda_i)\boldsymbol u_\lambda(Y_i,\boldsymbol \lambda_i)],
\\    
\boldsymbol\nabla_{jk} V_{\alpha}^I(Y_i|\boldsymbol{\theta},t_i, t_{i-1})
&= \left(\pdv{\boldsymbol\lambda_i}{\theta_j}{\theta_k}\right)^T(1+\alpha)[\boldsymbol C_{\alpha}(\boldsymbol\lambda_i)-f^\alpha(Y_i,\boldsymbol\lambda_i)\boldsymbol u_\lambda(Y_i,\boldsymbol\lambda_i)]
\\
& ~~+ (1+\alpha)\left(\pdv{\boldsymbol\lambda_i}{\theta_j}\right)^T\bigg[\boldsymbol C^{(1)}_{\alpha}(\boldsymbol\lambda_i)+(1+\alpha)\boldsymbol C^{(2)}_{\alpha}(\boldsymbol\lambda_i)-f^\alpha(Y_i,\boldsymbol\lambda_i)\boldsymbol \nabla_\lambda \boldsymbol u_\lambda(Y_i,\boldsymbol\lambda_i)
\\
& ~~~~~~~~~~~~~~~~~~~~~~~~~~~~~~~~~~~~~~~~~~~~~~
-\alpha f^\alpha(Y_i,\boldsymbol\lambda_i)\boldsymbol u_\lambda(Y_i,\boldsymbol\lambda_i) \boldsymbol u_\lambda(Y_i,\boldsymbol\lambda_i)^T\bigg] \left(\pdv{\boldsymbol\lambda_i}{\theta_k}\right),
\end{align*}
and  hence,
$$
E_{\boldsymbol\lambda_i}(\boldsymbol\nabla_{j}V_{\alpha}^I(Y_i|\boldsymbol{\theta},t_i, t_{i-1}))=0, ~~\mbox{ and }~~
\boldsymbol E_{\boldsymbol\lambda_i}(\boldsymbol\nabla_{jk}V_{\alpha}^I(Y_i|\boldsymbol{\theta},t_i, t_{i-1}))
=(1+\alpha) \left(\pdv{\boldsymbol\lambda_i}{\theta_j}\right)^T \boldsymbol C^{(2)}_{\alpha}(\boldsymbol\lambda_i)\left(\pdv{\boldsymbol\lambda_i}{\theta_k}\right).
$$
Now, if we denote by $\boldsymbol\Lambda_i$ the $\ell\times p$ matrix whose $j$th column is 
$\pdv{\boldsymbol\lambda_i}{\theta_j}$, we can see that
\[
\boldsymbol J^{(i)}(\boldsymbol{t})= (1+\alpha)\boldsymbol\Lambda_i^T \boldsymbol C^{(2)}_{\alpha}(\boldsymbol\lambda_i)\boldsymbol\Lambda_i,
~~~\mbox{ so that }~~~
\boldsymbol\Psi_n(\boldsymbol{t})= \frac{1}{n}\sum_{i=1}^n(1+\alpha)\boldsymbol\Lambda_i^T \boldsymbol C^{(2)}_{\alpha}(\boldsymbol\lambda_i)\boldsymbol\Lambda_i.
\]
Similarly, we can also obtain 
\begin{align*}
\boldsymbol \Omega_n(\boldsymbol{t})=& \frac{1}{n}\sum_{i=1}^n (1+\alpha)^2\boldsymbol\Lambda_i^T 
\left[\boldsymbol C^{(2)}_{2\alpha}(\boldsymbol\lambda_i)-\boldsymbol C_{\alpha}(\boldsymbol\lambda_i)\boldsymbol C_{\alpha}(\boldsymbol\lambda_i)^T
\right]\boldsymbol\Lambda_i.
\end{align*}

Before we write down the simplified sufficient conditions for this current special case, 
let us consider the following notation. For any real matrix $\boldsymbol{M}=((m_{ij}))$, 
we define its $\ell_2$-norm as 
$\|\boldsymbol M\|_2= \sup_{\boldsymbol x:\|\boldsymbol x\|_2\leq 1} \|\boldsymbol M \boldsymbol x\|_2$
and its $\ell_\infty$-norm as $\|\boldsymbol{M}\|_\infty = \sup_{i,j}|m_{ij}|$.
The $\ell_\infty$-norm of a vector is also defined similarly.

\bigskip\noindent
\textbf{Specific Assumptions for single family of increment distributions in IIP:} 
\setlist[condenum]{label=(SF\arabic*), ref=SF\arabic*, wide}
\begin{condenum}
    \item The parameter space ($\Lambda$) of the family of densities $f(\cdot;\boldsymbol\lambda)$ 
    is a compact subset of $\mathbb{R}^\ell$. 
    Further, the density $f(y;\boldsymbol\lambda)$ has its support independent of $\boldsymbol\lambda$, 
    and is bounded by a finite constant $M>0$ for all $\boldsymbol\lambda \in \Lambda$, and $y \in \mathcal{X}$. 
    The density is thrice differentiable with respect to $\boldsymbol\lambda$ having continuous derivatives. Further the integral $\int f^{1+\alpha}(y;\boldsymbol\lambda)dy$ can be differentiated thrice 
    with respect to $\boldsymbol\lambda$ for any $\alpha>0$ and the derivatives can be taken under the integral sign.
    \label{Single Family ass 1}
    
    \item $\sup_{i>1}\|\boldsymbol{\Lambda}_i\|_\infty= O(1)$ and $\sup_{i>1}\|\boldsymbol\nabla \boldsymbol{\Lambda}_i\|_\infty= O(1)$,
    and also $\sup_{n\geq 1}\|\boldsymbol\Omega_n^{-1/2}\|=O(1)$.
    \label{Single Family ass 2}
    
    \item Further if we define, for all $i=1, 2, \ldots$, 
    $$ 
    W_{2,i}(Y)= \|\boldsymbol u_\lambda(Y,\boldsymbol\lambda_i)\|_\infty+\|\boldsymbol\nabla_{\boldsymbol{\lambda}}\boldsymbol u_\lambda(Y,\boldsymbol\lambda_i)\|_\infty,+ \|\boldsymbol u_\lambda(Y,\boldsymbol\lambda_i)\boldsymbol u_\lambda(Y,\boldsymbol\lambda_i)^T\|_\infty,
    $$ 
then we assume
\begin{align*}
\lim_{N\to \infty} \sup_{n>1} \frac{1}{n}\sum_{i=1}^n P_{\boldsymbol\lambda_i}[W_{2,i}(Y)>N]= 0,
~~~\mbox{ and }~~
\lim_{N\to \infty} \sup_{n>1} \frac{1}{n}\sum_{i=1}^n E_{\boldsymbol\lambda_i}[W_{2,i}(Y)I(W_{2,i}(Y)>N)]= 0.
\end{align*}
Additionally, for every $\epsilon>0$, we also have
        \begin{align*}
&\lim_{n\to \infty}\frac{1}{n}\sum_{i=1}^n P_{\boldsymbol\lambda_i}\left[\|\boldsymbol u_\lambda(Y,\boldsymbol\lambda_i)\|_2>\epsilon n\right]= 0,
~~~ 
	\lim_{n\to \infty}\frac{1}{n}\sum_{i=1}^n E_{\boldsymbol\lambda_i}\left[\|\boldsymbol u_\lambda(Y,\boldsymbol\lambda_i)\|_2 I(\|\boldsymbol u_\lambda(Y,\boldsymbol\lambda_i)\|_2>\epsilon n)\right]= 0,  ~~\mbox{ and }~
\\   & \lim_{N\to \infty} \sup_{n>1} \frac{1}{n}\sum_{i=1}^n P_{\boldsymbol\lambda_i}\left[\|\boldsymbol u_\lambda(Y,\boldsymbol\lambda_i)\|_\infty>N\right]= 0,
~~~
\lim_{N\to \infty} \sup_{n>1} \frac{1}{n}\sum_{i=1}^n E_{\boldsymbol\lambda_i}\left[\|\boldsymbol u_\lambda(Y,\boldsymbol\lambda_i)\|_\infty I(\|\boldsymbol u_\lambda(Y,\boldsymbol\lambda_i)\|_\infty>N)\right]= 0.
    \end{align*}
\label{Single Family ass 3}
\end{condenum}

It is easy to see that, in this special case of  single family of increment distributions under the IIP models,
Assumption (\ref{Single Family ass 1}) implies the first three common Assumptions (\ref{Indep ass 1})--(\ref{Indep ass 3}).
Other two Assumptions (\ref{Single Family ass 2}) and (\ref{Single Family ass 3}), combined with (\ref{Single Family ass 1}), imply the seemingly difficulty Assumptions (\ref{Indep ass 6}) and (\ref{Indep Ass 7}), 
and finally as a result, the asymptotic properties of the MDPDEs follow as shown in the following theorem.

\bigskip
\begin{theorem}[Single Family of Distributions in IIP]\label{single increment family theorem}
Consider the discretely observed data at a pre-fixed time stamp vector $\boldsymbol{t}$ from an IIP model,
where the true increment distributions belong to one single family of distributions as mentioned above. 
Let $\alpha\geq 0$ is fixed. Then, under Assumptions (\ref{Single Family ass 1})--(\ref{Single Family ass 3}), (\ref{Indep ass 4}) and (\ref{Indep ass 5}) 
then the resulting MDPDE of the underlying parameter  satisfies the asymptotic properties 
as stated in Theorem \ref{main theorem}.
\end{theorem}

The above theorem considerably simplifies the required sufficient conditions from our general Theorem \ref{main theorem},
but Assumption (\ref{Single Family ass 3}) still looks a little bit complicated and hard to check.
However, it can be shown to hold for practically important subclasses of the IIPs as shown in the following corollaries. 

\bigskip
\begin{corollary}[Poisson process]
	\label{Poisson family theorem}
Consider the discretely observed data at pre-fixed time stamp vector $\boldsymbol{t}$ from an IIP model,
where the true increment distributions are Poisson with 
$f(y;\boldsymbol{\theta},t_i,t_{i-1})$ being the Poisson density with parameter  $\lambda_i= \Lambda(t_i)-\Lambda(t_{i-1})$;
here $\Lambda(s)= \Lambda(s,\boldsymbol{\theta})$ in the notation of Definition \ref{DEF:Poisson_proc}. 
Let $\alpha\geq 0$ is fixed. Then, Assumptions (\ref{Single Family ass 1})--(\ref{Single Family ass 2}) imply (\ref{Single Family ass 3}).
Therefore, the asymptotic properties of the MDPDEs, as stated in Theorem \ref{main theorem}, follow
under only the simpler Assumptions (\ref{Indep ass 4}), (\ref{Indep ass 5}) and 
(\ref{Single Family ass 1})--(\ref{Single Family ass 2}).
\end{corollary}

\bigskip
\begin{corollary}[Location-scale family for increment distributions in IIP]
	\label{location scale family theorem}
Consider the discretely observed data at pre-fixed time stamp vector $\boldsymbol{t}$ from an IIP model,
where the true increment distributions belong to one single location-scale family of distributions 
having support as the whole of $\mathbb{R}$ and densities of the form
$$
f(y;\boldsymbol\theta,t_i,t_{i-1})= \frac{1}{\sigma_i}f\left(\frac{y-\boldsymbol\mu_i}{\sigma_i}\right)$$ 
with $\boldsymbol\mu_i=\mu(\boldsymbol\theta,t_i,t_i-1)$ and 
$\sigma_i=\sigma(\boldsymbol\theta,t_i,t_{i-1})$ for some functions $\mu$ and $\sigma$
(in terms of the notation of this section $\boldsymbol \lambda_i=(\boldsymbol \mu_i,\sigma_i)^T$). 
Let $\alpha\geq 0$ is fixed. Then, whenever the relevant integrals exist, 
Assumptions (\ref{Single Family ass 1})--(\ref{Single Family ass 2}) imply (\ref{Single Family ass 3}).
Therefore, the asymptotic properties of the MDPDEs, as stated in Theorem \ref{main theorem}, follow
under only the simpler Assumptions (\ref{Indep ass 4}), (\ref{Indep ass 5}) and 
(\ref{Single Family ass 1})--(\ref{Single Family ass 2}).
\end{corollary}

\bigskip
\begin{corollary}[drifted Brownian Motion]
	\label{DB family theorem}
As a special case of Corollary \ref{location scale family theorem}, 
all the asymptotic results related to the MDPDEs hold for the drifted Brownian process 
defined in Section \ref{SEC:MDPDE_IIP} (where the increment density in normal)
with the relevant matrices (in its asymptotic variance) having the  simplified forms as given by 
\[
\boldsymbol\Psi_n(\boldsymbol{t})=\sum_{i=1}^n \frac{(2\pi)^{-\frac{\alpha}{2}}}{n(1+\alpha)^{1/2}\sigma_i^{2+\alpha}}\boldsymbol\Lambda_i^T \begin{bmatrix}
\frac{1}{1+\alpha} & 0\\
0 & \frac{3}{(1+\alpha)^2}-\frac{2}{1+\alpha}+1
\end{bmatrix}\boldsymbol\Lambda_i,
\]
and
\[
\quad \boldsymbol\Omega_n(\boldsymbol{t})= \sum_{i=1}^n \frac{(1+\alpha)^{3/2}(2\pi)^{-\frac{\alpha}{2}}}{n\sigma_i^{2+\alpha}}\boldsymbol\Lambda_i^T \begin{bmatrix}
\frac{1}{1+\alpha} & 0\\
0 & \frac{3}{(1+\alpha)^{2}}-\frac{2}{1+\alpha}+1-\frac{(2\pi)^-{\frac{\alpha}{2}}}{\sqrt{1+\alpha}}\left(\frac{1}{1+\alpha}-1\right)^2
\end{bmatrix}\boldsymbol\Lambda_i.
\]
\end{corollary}

Further (empirical) illustrations about the performances of the MDPDEs under the Poisson and drifted Brownian motion processes
will be presented in Section \ref{SEC:Examples}.

\subsection{Special Case II: m-dependent Stationary Markov Process}

We now consider a special subclass of the MPs, namely the m-dependent stationary MPs, and simplify Assumptions (\ref{MP ass 1})--(\ref{MP ass 3}).
This special subclass is defined based on the following two criteria. 
\begin{itemize}
	\item A stochastic process $\{X(t)\}_{t\geq 0}$  is said to be strictly stationary if 
$(X(t_i),X(t_{i+1}),\hdots,X(t_{i+k}))$ and $(X(t_j),X(t_{j+1}),\hdots,X(t_{j+k}))$ have the same distributions, for all $k\geq 1$ and $0<t_1<t_2<\cdots$.

	\item A stochastic process $\{X(t)\}_{t\geq 0}$  is said to be m-dependent if $\{X(t_i)\}_{i\leq k}$ is jointly-independent of $\{X(t_i)\}_{i\geq k+m+1}$, 
	for all $k\geq 1$ and $0<t_1<t_2<\cdots$.
\end{itemize}

Now, suppose that $\{X_i\}_{i\geq 0}$ is an m-dependent stationary process. 
Then, by definition, $\{(X_i,X_{i-1})\}_{i\geq 1}$ is (m+1)-dependent stationary stochastic process.
We can then replace Assumptions (\ref{MP ass 1})--(\ref{MP ass 3}) with the following one which are easier to check in practice.

\bigskip\noindent
\textbf{Specific Assumption for m-dependent Stationary Markov Process:} 
\setlist[condenum]{label=(MD\arabic*), ref=MD\arabic*, wide}
\begin{condenum}
	\item If we denote $V_1(\boldsymbol\theta)=V_\alpha^M(X_1,X_0|\theta,t_1,t_{0})$, then for every $j, k = 1, \ldots, p$, $E[|\nabla_k V_1(\boldsymbol\theta)|^3]$ 
	and $E[\nabla_{jk} V_1(\boldsymbol\theta)]^2$ exist finitely.
	
\end{condenum}
\smallskip

Note that, under this set-up, we have
$\boldsymbol\Omega(\boldsymbol{t})=E\left[\nabla V_1(\boldsymbol\theta)(\nabla V_1(\boldsymbol\theta))^T\right]+E\left[\sum_{i=2}^{m+1}\nabla V_1(\boldsymbol\theta)(\nabla V_i(\boldsymbol\theta))^T\right]$
and $\boldsymbol\Psi_n(\boldsymbol{t})=E\left[\nabla^2(V_1(\boldsymbol\theta))\right]$.
Under (MD1), the above expression makes sense and $\boldsymbol\Omega_n(\boldsymbol{t})\to \boldsymbol\Omega(\boldsymbol{t}) \text{ as } n\to \infty$.
Then, we can simplify our main theorem for the present case as follows.

\medskip
\begin{theorem}[m-dependent Stationary Markov Process]\label{m-dependent theorem}
Consider the discretely observed data at pre-fixed time stamp vector $\boldsymbol{t}$ from m-dependent stationary Markov process as mentioned above. 
Let $\alpha\geq 0$ is fixed. Then, under Assumptions (\ref{Indep ass 1})--(\ref{Indep ass 4}) and (MD1), 
 the resulting MDPDE of the underlying parameter  satisfies the asymptotic properties as stated in Theorem \ref{main theorem}.
\end{theorem}

The proof of Theorem \ref{m-dependent theorem} will follow in the same way the proof of our main Theorem \ref{main theorem} for general Markov processes
but by using the specific, simplified versions of WLLN and CLT for the present case of m-dependent stationary MPs.
The corresponding results, in consistent with our assumptions, are presented in the following lemma; 
see Appendix \ref{lemma proofs} for the proofs of the first two.

\begin{lemma}\label{m dependent wlln lemma}
Suppose $\{X_i\}_{i\geq 0}$ is a sequence of m-dependent strictly stationary random variables such that $\sigma^2=E(X_1^2)$ exists finitely.
Let $S_n$ be the partial sum up to $X_n$. Then, as $n\to \infty$, it follows that
$$\frac{S_n-E(S_n)}{n}\overset{P}{\to}0.$$
\end{lemma}

\begin{lemma}\label{m dependent lemma 2}
Suppose $X_1, X_2, \ldots$ be a stationary and m-dependent sequence of random variables such that $EX_1=0$ and $E|X_1|^3<\infty$. 
Let $S_n$ be the partial sum up to $n$ terms. Then, as $n\to \infty$, it follows that
\[
\frac{Var(S_n)}{n}\to A:=EX_1^2+2EX_1X_2+\ldots +2EX_1X_{m+1}.
\]
\end{lemma}

\begin{lemma}[\cite{hoeffding1948central}, Theorem 1, p.~206]\label{m dependent CLT}
Let $X_1,X_2,\ldots$ be a stationary and m-dependent sequence of random variables such that $EX_1=0$ and $E|X_1|^3<\infty$.
Let $S_n$ be the partial sum up to $n$ terms, and $A=EX_1^2+2EX_1X_2+\ldots +2EX_1X_{m+1}$. Then, as $n\to \infty$, it follows that
	\[
	\frac{1}{\sqrt{n}}S_n\overset{d}{\to}\mathcal{N}(0, A).
	\]
\end{lemma}

\biblio 

\section{Empirical Illustrations}
\label{SEC:Examples}
     

\subsection{Poisson Process}

\begin{figure}[h]
    \centering
    \includegraphics[scale=.9]{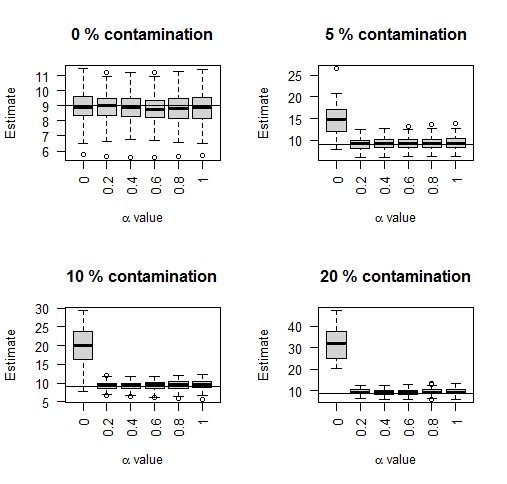}
    \caption{Boxplots of the MDPDEs for the Poisson process example}
    \label{box plot poisson}
\end{figure}

In line with our motivating example in Section \ref{SEC:Motivation}, let us first consider the Poisson process with intensity function $\lambda(t)=\frac{\theta}{2\sqrt{t}}$ for $t >0$, 
where $\theta$ is a single unknown parameter that we want to estimate. Hence, here our $i$-th increment distribution is given by 
$$X(t_i)-X(t_{i-1})\sim Poi(\theta(\sqrt{t_i} - \sqrt{t_{i-1}})).$$
We consider the time-stamp vector $\boldsymbol t=\mathbb{N}\cup\{0\}$. 
We fix our sample size to be $n=50$ and the true parameter value to be $\theta=9$. 
Additionally, to study the claimed robustness property, we contaminate the simulated (pure) data via the following scheme: 
at $\delta\%$  level of contamination, our increment distribution for all $i$ is given by 
\[
X(t_i)-X(t_{i-1})\sim \frac{(100-\delta)}{100}Poi(\theta(\sqrt{t_i} - \sqrt{t_{i-1}}))+\frac{\delta}{100}Poi(18).
\]
We compute the MDPDE for different tuning parameters based on the pure data  (0\% contamination) as well as the contaminated data for 5\%, 10\% and 20\% contamination proportion ($\delta$).
We replicate this process 200 times, and report the average values of the resulting MDPDEs, along with their empirical mean squared errors (MSEs),  in Table \ref{mse_poisson};
the box-plots of the MDPDEs obtained for these 200 replications are also shown in Figure \ref{box plot poisson}.

\begin{table}[ht!]
    \centering
\caption{The MDPDEs, along with their MSEs (within parentheses) for the Poisson process example}
    \begin{tabular}{|l|c|c|c|c|}
    \hline
    \hline
    contamination & 0\% & 5\% & 10 \% & 20 \%\\
    \hline
    \hline
   $\alpha =0$ (MLE) & 8.958(1.1)   & 14.936(7.041) & 20.07(12.248) & 32.038(24.087) \\
   $\alpha =0.2$ & 8.921(1.056) & 9.212(1.38)   & 9.317(1.186)  & 9.546(1.519)   \\
   $\alpha =0.4$ & 8.882(1.054) & 9.238(1.462)  & 9.263(1.269)  & 9.404(1.489)   \\
   $\alpha =0.6$ & 8.857(1.074) & 9.3(1.559)    & 9.319(1.378)  & 9.522(1.604)   \\
   $\alpha =0.8$ & 8.845(1.102) & 9.365(1.652)  & 9.392(1.483)  & 9.672(1.757)   \\
   $\alpha =1$ & 8.842(1.137) & 9.426(1.734)  & 9.47(1.584)   & 9.825(1.922)  \\
     \hline
    \end{tabular}
    \label{mse_poisson}
\end{table}

It can clearly seen that the  MLE (at $\alpha = 0$) gets more and more away from the true parameter value (of 9) as the contamination level increases; 
also its MSE increases steeply with increasing contamination. For non-zero contamination levels, the bias of MLE is clearly visible in the figures, which further increases for higher contamination proportion.
However,  for our proposed MDPDEs with $\alpha>0$,  increases in its bias and MSE are substantially lower compared to that of the MLE. 
As $\alpha$ increases, the bias and MSE of the corresponding MDPDEs become more stable. 
For no contamination,  we see that the MLE and our proposed MDPDEs have similar bias and MSE indicating no significant loss for the MDPDEs even under pure data.
Further, the MSEs attain the minimum at a suitable $\alpha>0$ for any non-zero contamination level.

\subsection{Drifted Brownian Motion}

\begin{figure}
  \centering
  \includegraphics[scale=0.8]{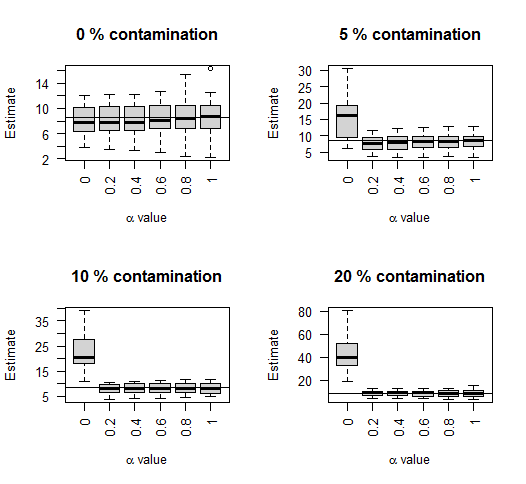}
  \caption{Boxplot of MDPDEs of $\theta_1$ for the drifted Brownian model example}
  \label{box plot param1 brow}
\end{figure}
\begin{figure}
  \centering
  \includegraphics[scale=0.8]{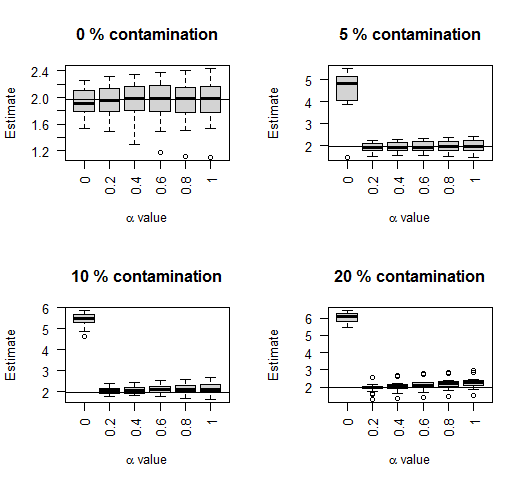}
  \caption{Boxplot of MDPDEs of $\theta_2$ for the drifted Brownian model example}
  \label{box plot param2 brow}
\end{figure}

Let us now consider a drifted Brownian motion process with the mean and sigma functions being given by 
\begin{align*}
\mu(t,\boldsymbol\theta)=&\theta_1 \sqrt{t}, ~~~~\mbox{ and }~~~\sigma(\boldsymbol\theta)=e^{\theta_2},
\end{align*}
where the parameter of interest $\boldsymbol\theta=(\theta_1,\theta_2)$ is a vector. Then, our $i$-th increment distribution is given by 
$$
f(s;\boldsymbol\theta,t_i,t_{i-1})=\phi\left(\frac{s-\mu(t_i,\boldsymbol\theta)+\mu(t_{i-1},\boldsymbol\theta)}{\sigma(\boldsymbol\theta)(t_i-t_{i-1})}\right),
$$
where $\phi(\cdot)$ represents standard normal density function. 
We consider the time-stamp vector $\boldsymbol t=\mathbb{N}\cup\{0\}$. 
Now we fix our sample size to be $n=50$ and the true parameter value to be $\boldsymbol \theta_g=(8.5,1.97)$.  
Further, to examine robustness, we again contaminate the simulated (pure) sample data as follows: 
at $\delta\%$  level of contamination,  the $i$-th increment distribution is considered to be 
\[
\frac{(100-\delta)}{100}\phi\left(\frac{s-\mu(t_i,\boldsymbol\theta)+\mu(t_{i-1},\boldsymbol\theta)}{\sigma(\boldsymbol\theta)(t_i-t_{i-1})}\right)+\frac{\delta}{100}\phi\left(\frac{s-50}{1.3}\right),
\]
for all $i=2, \ldots, n$.
By replicating the process 200 times, we report the average values of the MDPDEs, along with their empirical MSEs, in Table \ref{mse_param2_brow};
the box-plots of the resulting MDPDEs of $\theta_1$ and $\theta_2$ are also presented separately in Figures \ref{box plot param1 brow} and figure \ref{box plot param2 brow}, respectively. 

\begin{table}[ht!]
    \centering
\caption{The MDPDEs of $\theta_1$, along with their MSEs (within parentheses) for the drifted Brownian model example}
    \begin{tabular}{|l|c|c|c|c|}
    \hline
    \hline
    contamination & 0\% & 5\% & 10 \% & 20 \%\\
    \hline
    \hline
    \multicolumn{5}{|l|}{\textbf{For parameter $\theta_1$}}\\
    $\alpha=0$ (MLE) & 8.151(2.267) & 15.661(9.724) & 22.853(16.049) & 42.348(36.883) \\
    $\alpha=0.2$  & 8.252(2.292) & 7.804(2.424)  & 7.94(2.007)    & 8.55(2.559)    \\
    $\alpha=0.4$  & 8.345(2.352) & 7.958(2.467)  & 7.924(2.106)   & 8.646(2.671)   \\
    $\alpha=0.6$  & 8.483(2.73)  & 8.11(2.455)   & 7.925(2.185)   & 8.724(2.776)   \\
    $\alpha=0.8$  & 8.622(3.092) & 8.228(2.488)  & 7.918(2.221)   & 8.818(2.851)   \\
    $\alpha=1$  & 8.688(3.254) & 8.446(2.502)  & 7.941(2.207)   & 9.007(3.34)  \\
        \hline
    \hline
    \multicolumn{5}{|l|}{\textbf{For parameter $\theta_2$}}\\
    $\alpha=0$ & 1.941(0.206) & 4.565(2.749) & 5.408(3.458) & 6.031(4.071) \\
    $\alpha=0.2$ & 1.948(0.236) & 1.193(0.216) & 2.044(0.19)  & 1.954(0.287) \\
    $\alpha=0.4$ & 1.948(0.271) & 1.93(0.234)  & 2.079(0,21)  & 2.036(0.3)   \\
    $\alpha=0.6$ & 1.941(0.291) & 1.948(0.257) & 2.11(0.24)   & 2.116(0.332) \\
    $\alpha=0.8$ & 1.936(0.301) & 1.965(0.277) & 2.134(0.271) & 2.182(0.371) \\
    $\alpha=1$ & 1.937(0.305) & 1.974(0.297) & 2.158(0.299) & 2.231(0.404)\\
    \hline
    \end{tabular}
\label{mse_param2_brow}
\end{table}

We can again see that the MLE (at $\alpha = 0$) is extremely non-robust having significantly increased values of both bias and MSE as contamination level increases.
For the proposed MDPDEs with $\alpha>0$, the changes in the estimates as well as their bias and MSEs are substantially lower compared to those of the MLE under contaminated data;
these stability further increases as $\alpha$ increases indicating the increased robustness of the corresponding MDPDEs.
Further, under pure data, the bias and MSE are minimum for the MLE as expected, but the corresponding loss (in terms of either the bias or the MSE)  are not quite significant for the MDPDEs with $\alpha>0$.

\subsection{AR(1) Process}

\begin{figure}[!h]
  \centering
  \includegraphics[scale=.9]{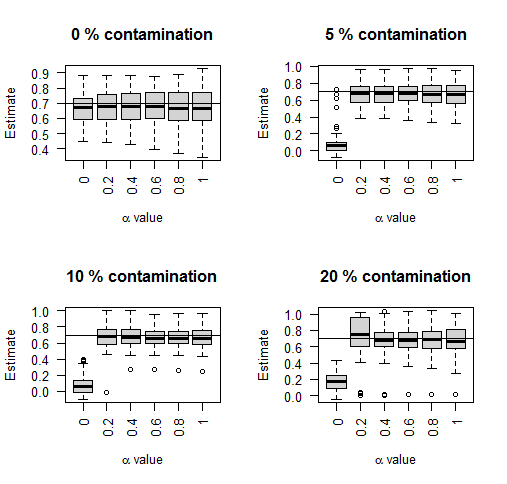}
  \caption{Boxplots of the MDPDEs for the AR(1) process example}
  \label{box plot AR 1 process}
\end{figure}

As our final illustration, let us note that auto-regressive (AR) time series can also be considered as a special case of our general set-up (although it was previously studied separately).
So, let us now illustrate the finite sample performance of our proposed MDPDE under a special case of the AR time series, 
namely the AR(1) process defined on a pre-fixed countable index set $T\subseteq \mathbb{R}$ as
$$
X(t_i)=\theta X(t_{i-1})+\epsilon_{i}, \quad i\geq 1,  \quad \epsilon_1, \epsilon_2, \ldots \mathop{\sim}^{\mbox{IID}} N(0,1),
$$
where $\theta$ is again a single parameter of interest.  
Hence, here our $i$-th conditional distribution is given by 
$$
X(t_i)|X(t_{i-1})\sim \mathcal{N}(\theta X(t_{i-1}),1).
$$
We consider the time-stamp vector $\boldsymbol t=T$, fix our data size to $n=50$ and the true parameter value to be $\theta_g=0.7$. 
Additionally, we consider a contamination scheme as follows: at $\delta\%$  level of contamination, for every $i$,   the $i$-th conditional distribution is taken as, 
\[
X(t_i)|X(t_{i-1})\sim \frac{100-\delta}{100}\mathcal{N}(\theta X(t_{i-1}),1)+\frac{\delta}{100  }\mathcal{N}(25,1.3).
\]
We replicate this simulation exercise 200 times to compute the empirical MSEs of the MDPDEs of $\theta$, which are reported in Table \ref{mse_ar1}, 
along with the average values of the resulting MDPDEs; their box-plots are also presented in Figure \ref{box plot AR 1 process} for different $\alpha$ 
under pure data as well as under different levels of contamination.

\begin{table}[ht!]
    \centering
\caption{The MDPDEs, along with their MSEs (within parentheses) for the AR(1) process example}
	\begin{tabular}{|l|c|c|c|c|}
    \hline
    \hline
    contamination & 0\% & 5\% & 10 \% & 20 \%\\
    \hline
    \hline
    $\alpha=0$ (MLE) & 0.667(0.103) & 0.102(0.623) & 0.082(0.629) & 0.169(0.543) \\
    $\alpha=0.2$ & 0.67(0.107)  & 0.673(0.136) & 0.681(0.169) & 0734(0.258)  \\
    $\alpha=0.4$ & 0.671(0.113) & 0.672(0.138) & 0.685(0.15)  & 0.684(0.212) \\
    $\alpha=0.6$ & 0.671(0.12)  & 0.666(0.138) & 0.662(0.129) & 0.673(0.206) \\
    $\alpha=0.8$ & 0.67(0.128)  & 0.664(0.143) & 0.66(0.133)  & 0.675(0.21)  \\
    $\alpha=1$ & 0.669(0.138) & 0.661(0.146) & 0.659(0.139) & 0.68(0.201) 
    \\
    \hline
    \end{tabular}
\label{mse_ar1}
\end{table}

The simulation findings are again similar to the previous two cases. The MLE is seen to be highly non-robust yielding greater bias and MSE under contamination, 
but has the minimum MSE under pure data. Our proposed MDPDEs perform significantly robustly, proving much stable bias and variances, under data contamination 
with only a slight (not so significant) loss under pure data. The robustness of the MDPDEs further increase as $\alpha>0$ increase.

\biblio 


\newpage
\section{A Real Data Application}
\label{SEC:Data}
    


Let us now apply our proposed MDPDE for the analysis of an interesting dataset, namely the von Bortkeiwicz Prussian horse-kicking data \cite{von1898gesetz}.
These data were collected over a $20$ year period $(1875-1894)$ from $14$ different Prussian cavalry corps 
and contain the number of soldiers that died by horse kick for each division. 
The dataset is available in \texttt{R} under the package \texttt{vcd} and listed on \texttt{Rdatasets} by the name \texttt{prussian}.

This dataset is particularly popular and exemplifies a classic Poisson process, because death by horse-kicking is a rare, random event and 
a horse-kicking event in one corps is not reliant on the previous occurrence in that or any other corps. 
More specifically, a Poisson distribution here models the probability of the number of horse-kicking deaths for given corps for certain period of time.
Thus, we may model the data on the total number of deaths for each cavalry corps by a homogeneous Poisson process having intensity $\lambda(t)=\lambda$ for all $t>0$. 
Note that, as a consequence, for a particular corp, the death counts for every year are then IID realization from Poisson Distribution which is well verified in the literature. 
For illustration, in the following, we will only report the analysis and results for the first cavalry corp which is the guard corp (marked as ``G" in the dataset)l;
the results for other corps are verified to have similar nature and hence theyr are not reported for brevity. 

Note that, for the corp ``G" (as in with other corps), there are $n=20$ data points corresponding to each year from $1875-1984$;
we relabel these time indices to $\{1,2,\hdots ,20\}$ in our modeling purposes. Since there is no inherent clear outlies in the data, 
to illustrate the robustness of the proposed MDPDEs, we add artificial contamination as follows: 
at $\delta\%$ level of contamination, for each data point, we replace the original observation by a random observation from Poisson($18$) distribution with probability $\frac{\delta}{100}$.
The proposed MDPDEs (and the MLE) of the parameter ($\lambda$) of the fitted homogeneous Poisson process are reported in Table \ref{prussian estimators} for different values of the tuning parameter $\alpha$ 
and different levels of contamination.

\begin{table}[!ht]
	\centering
\caption{The MDPDEs of the parameter ($\lambda$) of the homogeneous Poisson process model for the Prussian Horse-kicking data (without and) with different levels of artificial contamination}
		\begin{tabular}{|l|c|c|c|c|}
			\hline \hline
			contamination  & 0\% (Original) & 5\%  & 10\% & 20\% \\
			\hline \hline
			$\alpha=0$ (MLE) & 0.8 & 1.742 & 2.47 & 3.423\\
			$\alpha=0.2$ & 0.808 & 0.825 & 0.836 & 0.86\\
			$\alpha=0.4$ & 0.810 & 0.83 & 0.843 & 0.874\\
			$\alpha=0.6$ & 0.809 & 0.834 & 0.854 & 0.898\\
			$\alpha=0.8$ & 0.806 & 0.837 & 0.862 & 0.918\\
			$\alpha=1$ & 0.804 & 0.839 & 0.869 & 0.935\\
			\hline
	\end{tabular}
\label{prussian estimators}
\end{table}

Note that the MDPDEs are quite close to the MLE for the original data as there is no contamination. 
However, after adding artificial contamination, the MLE increases rapidly to some other values which clearly do not fit the majority of the data well.
However, the proposed MDPDE with $\alpha>0$ remains much more stable (close to their original values) even under heavy contamination of 20\% 
indicating their suitability to generate correct inference even under data contamination.

\biblio 

\section{Conclusion}
\label{SEC:Conclusion}

In this paper we have discussed a robust efficient estimator, namely the MDPDEs, for the independent increment and the Markov processes. 
We have established asymptotic properties (consistency and asymptotic normality results) for the resulting MDPDEs under such general set-ups of discretely observed stochastic processes 
and simplified the required conditions for several important special model subclasses. Their finite-sample properties are illustrated through extensive simulation study and an interesting real data analysis. 
The claimed robustness of the MDPDEs under the IIPs are also justified theoretically through the influence function analysis in Appendix \ref{APP:A}.

This work opens up a new direction of further research on the robust and efficient parametric statistical inference for more general stochastic processes. 
An immediate future work would be to extend the MDPDEs for general stochastic process instead of restricting to specific dependence structure such as in Independent Increment Process or Markov Process. 
One possible approach could be to extend the conditional density based definition from the Markov process set-up to the general setup, 
through using the conditional density of $X(t_i)$ given $X(t_{i-1}),\hdots,X(t_0)$ in place of the density  of $X(t_i)$ given $X(t_{i-1})$
and minimizing the average DPD measures between the model and empirical estimates of these conditional densities. 
Although implementation would be rather straightforward, the derivation of the distributional results under such general set-up would be extremely difficult and non-trivial;
we hope to develop further on this idea in our future research works. 
Further research would also be required in order to define and study the theoretical robustness measure for the MDPDE under the Markov process or then for the more general stochastic processes.

\biblio 

\bigskip\bigskip
\begin{appendices}

\counterwithin{table}{section}
\counterwithin{figure}{section}


\section{Theoretical Robustness of the MDPDEs under the IIP Models}
\label{APP:A}

Let us consider the notation of Section \ref{SEC:MDPDE_IIP} and let $G(\cdot;t_i,t_{i-1})$ denote the true distribution for the datum $Y_i$ having density $g(\cdot;t_i,t_{i-1})$ for all $i$.
We can define the minimum DPD functional $\boldsymbol{T}_\alpha(G(\cdot;t_1,t_0),\hdots ,G(\cdot;t_n,t_{n-1}))$ for discrete data (with finite sample size $n$) observed from an IIP as the minimizer of 
\begin{equation}\label{IIP stat functional}
 \frac{1}{n}\sum_{i=1}^n d_\alpha(g(\cdot;t_i,t_{i-1}),f_i(\cdot;\boldsymbol\theta,t_i,t_{i-1})),
\end{equation}
with respect to $\boldsymbol{\theta}\in\Theta$. Note that, as in the definition of MDPDE, $\boldsymbol{T}_\alpha(G(\cdot;t_1,t_0),\hdots ,G(\cdot;t_n,t_{n-1}))$ 
can equivalently be defined as the minimizer of the simpler objective function 
\begin{equation*}
    \sum_{i=1}^n H^{(i)}_\alpha (\boldsymbol\theta)=\sum_{i=1}^n\Bigg[\int f^{1+\alpha}(y;\boldsymbol\theta,t_i,t_{i-1})dy-\left(1+\frac{1}{\alpha}\right)\int f^{\alpha}(y;\boldsymbol\theta,t_i,t_{i-1})dG(y;t_i,t_{i-1})\Bigg].
\end{equation*}
Under appropriate differentiability conditions, it leads to the estimating equation
\begin{equation*}
\begin{split}
   \sum_{i=1}^n\bigg[\int f^{1+\alpha}(y;\boldsymbol\theta,t_i,t_{i-1})& \boldsymbol u(y;\boldsymbol\theta,t_i,t_{i-1})dy\\
   &-\int f^{\alpha}(y;\boldsymbol\theta,t_i,t_{i-1})\boldsymbol u(y;\boldsymbol\theta,t_i,t_{i-1})g(y;t_i,t_{i-1})dy\bigg]=\boldsymbol{0}.
   \end{split}
\end{equation*}

To derive the Influence function (IF) for IIP set-up, we will follow the approach used by \cite{huber1983minimax} in the context of the influence function for the non-IID fixed-carriers linear models. 
We consider the contaminated density $g_{i,\epsilon}=(1-\epsilon)g(\cdot;t_i,t_{i-1})+\epsilon \delta_{r_i}$ where $\delta_{r_i}$ is the degenerate distribution at the the point of the contamination point $r_i$ 
for $i=1,2,\hdots,n$. Now, since the setup here is similar to that one of non-homogeneous independent setup, 
we set $\boldsymbol\theta=\boldsymbol{T}_\alpha(G(\cdot;t_1,t_0),\hdots ,G(\cdot;t_n,t_{n-1}))$ to follow the process of \cite{ghosh2013robust} 
and our required IF of the MDPDE under the present IIP set-up come out to be
\begin{equation*}
IF(r_1,r_2,\hdots; r_n,G(\cdot,t_1,t_0),\hdots,G(\cdot,t_n,t_{n-1}))
=\boldsymbol\Psi_{n}^{-1}\frac{1}{n}\sum_{i=1}^n\bigg[f^\alpha (r_i;\boldsymbol\theta,t_i,t_{i-1}) \boldsymbol u(r_i;\boldsymbol\theta,t_i,t_{i-1})-\boldsymbol\zeta_i\bigg],
\end{equation*}
where $\boldsymbol\zeta_i=\int f(y;\boldsymbol\theta,t_i,t_{i-1}) \boldsymbol u(y;\boldsymbol\theta,t_i,t_{i-1})g(y;\boldsymbol\theta,t_i,t_{i-1})dy$.

In particular, when the increment distributions come from a single increment distribution family, as per our notation in Corollary \ref{single increment family theorem}, 
This IF can simply be expressed as
\begin{equation*}
    IF(r_1,r_2,\hdots,r_n; G(\cdot,t_1,t_0),\hdots,G(\cdot,t_n,t_{n-1}))
    =\boldsymbol\Psi_{n}^{-1}\frac{1}{n}\sum_{i=1}^n\bigg[f(r_i;\boldsymbol\lambda_i)^\alpha \boldsymbol u(r_i;\boldsymbol\lambda_i)-\boldsymbol\zeta_i\bigg],
\end{equation*}
with $\boldsymbol\zeta_i=\int f(y;\boldsymbol\lambda_i) \boldsymbol u(y;\boldsymbol\lambda_i)g(y;\boldsymbol\lambda_i)dy$.
In the following, we illustrate the form and the nature of this IF for the examples of Poisson process and the drifted Brownian motion to theoretically justify the claimed robustness of the MDPDEs for these set-ups.

\subsection{Example 1: Poisson Process}

Let us consider the Poisson process model and assume that the true distributions belong the same Poisson family, i.e., $g(\cdot;t_i,t_{i-1})$ is the Poisson($\lambda_i$) density. 
In this case, we can easily derive that $\boldsymbol u(y,\lambda_i)=\boldsymbol\Lambda_i^T\frac{y-\lambda_i}{\lambda_i}$, and thus, 
\begin{equation*}
	\boldsymbol\psi_n=\frac{1}{n}\sum_{i=1}^n (1+\alpha)\boldsymbol\Lambda_i^T \boldsymbol C_{\alpha}^{(2)}(\lambda_i)\boldsymbol\Lambda_i,
	~~~~\mbox{ and }~~
	\boldsymbol\zeta_i=\sum_{X=1}^\infty\boldsymbol\Lambda_i^T \bigg(e^{-\lambda_i} \frac{\lambda_i^X}{X!}\bigg)^2\frac{X-\lambda_i}{\lambda_i}.
\end{equation*}
Hence, using our general form of influence function for IIP process, we have
\begin{equation*}
	IF(r_1,r_2,\hdots,r_n; G(\cdot,t_1,t_0),\hdots,G(\cdot,t_n,t_{n-1}))
	=\boldsymbol\Psi_{n}^{-1}\frac{1}{n}\sum_{i=1}^n\bigg[\boldsymbol\Lambda_i^T e^{-\alpha\lambda_i}\frac{\lambda_i^{\alpha r_i}}{(r_i!)^\alpha}\frac{r_i-\lambda_i}{\lambda_i}-\boldsymbol\zeta_i\bigg].
\end{equation*}
A close form precise expression of $\boldsymbol\Psi_n^{-1}$ or of $\boldsymbol\zeta_i$ or of $\boldsymbol C_{\alpha}^{(2)}(\lambda_i)$ is difficult to get without imposing any further structure;
we illustrate it for a specific example below.

\begin{figure}[!h]
	\centering
	\includegraphics[scale=0.6]{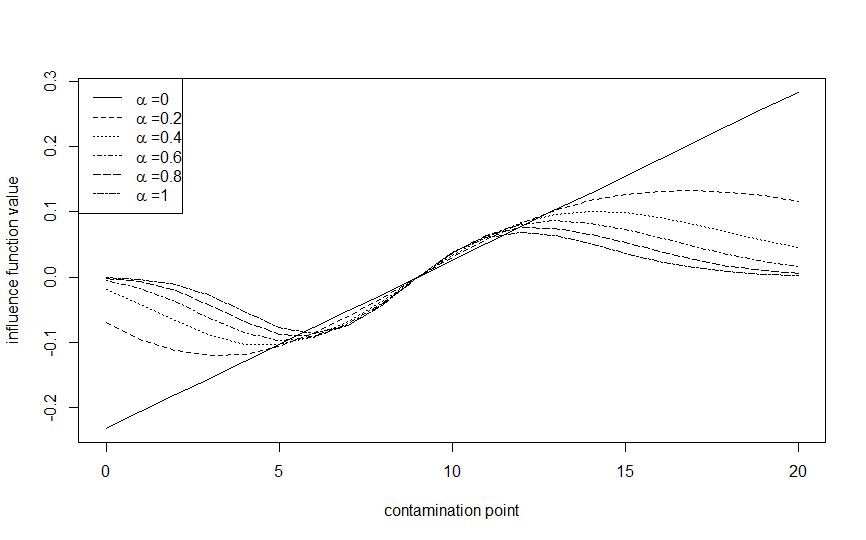}
	\caption{Influence Function plot for Poisson Process for $\theta=9$}
	\label{poisson influence}
\end{figure}

In consistent with our previous illustrations, e.g.,  as in Section \ref{SEC:Examples}.1, let us consider the Poisson process with the intensity function $\lambda(t)=\frac{\theta}{2\sqrt{t}}$ 
and a time stamp vector as $\boldsymbol{t}=(0,1,2,3,\hdots,50)$. Here $\theta$ is an uni-dimensional parameter and $\lambda_i=\theta(\sqrt{ t_i}-\sqrt{ t_{i-1}})$. Thus, we have 
\begin{equation*}
	\psi_n=\frac{1}{n}\sum_{i=1}^n  (1+\alpha)C_{\alpha}^{(2)}(\lambda_i)(\sqrt{ t_i}-\sqrt{ t_{i-1}})^2,
	~~~~~\mbox{ where }~~C_{\alpha}^{(2)}(\lambda_i)=\sum_{k\in \mathcal{N}}\frac{e^{-\lambda_i(1+\alpha)}\lambda_i^{k(1+\alpha)}(k-\lambda_i)^2}{(k!)^{1+\alpha}\lambda_i^2}.
\end{equation*}
Using these simplified formulas, we can then numerically compute the IF for different $\alpha$, which are plotted in Figure \ref{poisson influence} over the contamination points $r_1=r_2=\cdot=r_n$.
It can be seen that, in terms of the IF analysis, the effect of contamination is linearly increasing at $\alpha=0$, i.e., for the MLE; 
this unbounded nature of the IDF theoretically proves the non-robust nature of the MLE. 
But, as $\alpha$ increases towards $1$, the robustness of our proposed MDPDEs is more visible via their bounded IFs. 
The absolute value of the IF dips down significantly as we increase contamination point for high values of $\alpha$. 
Further, the maximum of these absolute IF values decreases as $\alpha$ increases indicating, theoretically, the increasing robustness of the corresponding MDPDEs,
in consistent with our earlier empirical illustrations. 

\par


\subsection{Example 2: Drifted Brownian Motion}

Let us now also derive the IF for the drifted Brownian motion assuming that the true distributions are coming from the same normal family, i.e., $g(\cdot;t_i,t_{i-1})$ is an univariate Normal($\mu_i,\sigma_i$) density. 
We have already derived the form of $\boldsymbol\Psi_n(\boldsymbol{t})$ in Corollary \ref{DB family theorem} with the notation $\mu_i=\mu(t_i)-\mu(t_{i-1})$ and $\sigma_i=\sigma(t_i-t_{i-1})$. 
Following calculations in Appendix \ref{proof of DB family theorem}, we also have 
\begin{equation*}
	\boldsymbol u(y,\boldsymbol\lambda_i)=\boldsymbol\Lambda_i^T\bigg(\frac{y-\mu_i}{\sigma_i^2},-\frac{1}{\sigma_i}+\frac{(y-\mu_i)^2}{\sigma_i^3}\bigg),
~~~~\mbox{ and }~~
	\boldsymbol\zeta_i=\boldsymbol\Lambda_i^T\frac{(2\pi)^{-\frac{1}{2}}}{\sqrt{2}\sigma_i^2}\times \left(0,-\frac{1}{2}\right).
\end{equation*}
Hence, we have the required IF as given by 
\begin{align*}
	&IF(r_1,r_2,\hdots,r_n,G(\cdot,t_1,t_0),\hdots,G(\cdot,t_n,t_{n-1}))\\
	& ~~~~~~~~~~~
	= \boldsymbol\Psi_{n}^{-1}\frac{1}{n}\sum_{i=1}^n\boldsymbol\Lambda_i^T\bigg(\frac{1}{(2\pi)^{\alpha/2}\sigma_i^\alpha}
	 e^-{\frac{\alpha(r_i-\mu_i)^2}{2\sigma_i^2}} \begin{bmatrix}
		\frac{y-\mu_i}{\sigma_i^2}\\-\frac{1}{\sigma_i}+\frac{(y-\mu_i)^2}{\sigma_i^3}
	\end{bmatrix}-\frac{(2\pi)^{-\frac{1}{2}}}{\sqrt{2}\sigma_i^2}\begin{bmatrix}0\\ -\frac{1}{2}\end{bmatrix}	
\bigg).
\end{align*}

\begin{figure}[!h]
	\centering
	\includegraphics[scale=0.6]{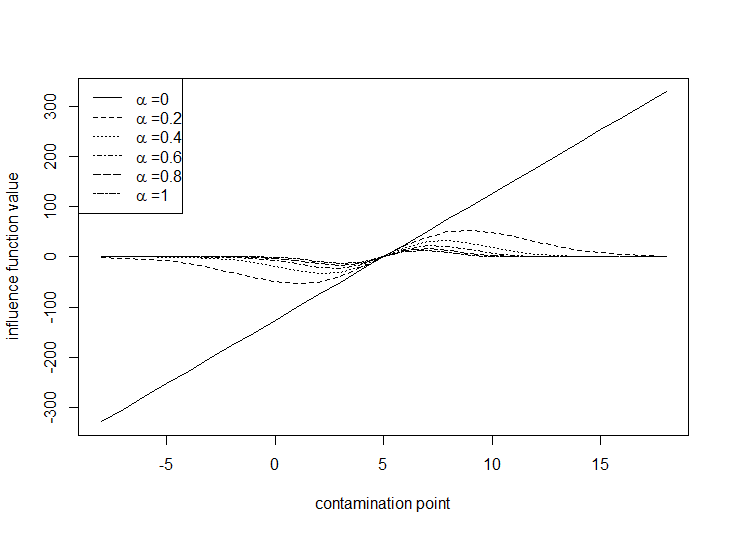}
	\caption{Influence Function plot for Drifted Brownian Motion for $\theta=5$}
	\label{brownian influence}
\end{figure}

To get a better visualization of the above IF, let us consider the an example with a specific mean and sigma function as given by  
$\mu(t;\theta)=\theta\sqrt{t}$ and $\sigma(\theta)=3$ (constant), 
along with the where time stamp vector being $\boldsymbol{t}=(0,1,2,3,\hdots,50)$.
Note that, clearly $\theta$ is again an uni-dimensional parameter and  $\boldsymbol\lambda_i=(\mu_i,\sigma_i)^T=(\theta(\sqrt{t_{i}}-\sqrt{t_{i-1}}),3)^T$. 
Thus, we get 
\begin{equation*}
	\boldsymbol\Psi_n(\boldsymbol t)=\frac{1}{n}\sum_{i=1}^n\frac{(2\pi)^{-\frac{\alpha}{2}}}{n(1+\alpha)^{1/2}\sigma_i^{2+\alpha}} (\sqrt{t_i}-\sqrt{t_{i-1}})^2.
\end{equation*}
We again numerically compute the IF of this particular example of drifted Brownian motion, using the simplified formulas, at various contamination points $r_1=\cdots=r_n$,
which is presented in Figure \ref{brownian influence} for different values of $\alpha$. 
The nature of these IFs are again the same as in the case of Poisson process example --- the unbounded IF of the MLE (at $\alpha=0$) theoretically justify its non-robust nature 
and the bounded redescending nature of the IFs of the MDPDEs with $\alpha>0$ further justify the claimed increasing robustness of the proposed MDPDEs with increasing $\alpha$. 
These are again in line with our empirical findings from the simulation study illustrating the concurrence of the numerical and theoretical results derived in the paper. 

\section{Proofs of additional theorems and corollaries}
\label{APP:B}

\subsection{Proof of Theorem \ref{single increment family theorem}}
We will need the following lemma to prove the theorem.

\begin{lemma}\label{single increment family lemma}
	Suppose that $A$ and $B$ are two non-negative random variables such that $A\leq B$ a.e.. 
	Then, for any $N>0$, we have
	\[
	AI(A>N)\leq BI(B>N) \quad \text{ a.e.}.
	\]
\end{lemma}
\begin{proof}
Suppose $\omega \in \{A>N\}$. Since, $A\leq B$ we also have $\omega \in \{B>N\}$. Hence, $\{A>N\}\subseteq \{B>N\}.$
Also, since $A$ is non-negative random variable, we have	$AI(A>N)\leq AI(B>N)\text{ a.e.}.$
Now, again using $A\leq B\text{ a.e.}$, we will have $AI(A>N)\leq BI(B>N) \text{ a.e.}.$	
Hence, proved.
\end{proof}

\medskip
\noindent
\textbf{Proof of the theorem:}\\
Remember, we are working under the setup that the increment distributions come from a single family of distribution. Also, we have assumed that the true distributions belong to the model family. 
It is easy to observe that (\ref{Single Family ass 1}) imply (\ref{Indep ass 1})--(\ref{Indep ass 3}). 
Thus, as per theorem \ref{main theorem}, it is enough to show that (\ref{Single Family ass 1})--(\ref{Single Family ass 3}) implies (\ref{Indep ass 6}) and (\ref{Indep Ass 7}).

\textit{First Condition of (\ref{Indep ass 6}):}
Now, first recall that 
\begin{align*}
    \boldsymbol\nabla_j V_i(X;\boldsymbol\theta,\boldsymbol{t})=&(1+\alpha)\left(\pdv{\boldsymbol\lambda_i}{\theta_j}\right)^T\bigg[\boldsymbol C_\alpha(\boldsymbol\lambda_i)-f^\alpha(X,\boldsymbol\lambda_i)\boldsymbol u_\lambda(X,\boldsymbol\lambda_i)\bigg]\\
    =&(1+\alpha)\sum_{m=1}^k\left(\pdv{\boldsymbol\lambda_i}{\theta_j}\right)_m\bigg[(\boldsymbol C_\alpha(\boldsymbol\lambda_i))_m-f^\alpha(X,\boldsymbol\lambda_i)(\boldsymbol u_\lambda(X,\boldsymbol\lambda_i))_m\bigg].
\end{align*}
Thus, we have following inequalities
\begin{align*}
    |\boldsymbol\nabla_j V_i(X;\boldsymbol\theta,\boldsymbol{t})|\leq & (1+\alpha)\sum_{m=1}^k\bigg|\left(\pdv{\boldsymbol\lambda_i}{\theta_j}\right)_m\bigg|\bigg[\bigg|(\boldsymbol C_\alpha(\boldsymbol\lambda_i))_m\bigg|+f^\alpha(X,\boldsymbol\lambda_i)\bigg|(\boldsymbol u_\lambda(X,\boldsymbol\lambda_i))_m\bigg|\bigg]\\
    \leq & (1+\alpha)\sum_{m=1}^k\bigg|\left(\pdv{\boldsymbol\lambda_i}{\theta_j}\right)_m\bigg|\bigg[\bigg|(\boldsymbol C_\alpha(\boldsymbol\lambda_i))_m\bigg|+M^\alpha\bigg|(\boldsymbol u_\lambda(X,\boldsymbol\lambda_i))_m\bigg|\bigg] \quad \text{[using (\ref{Single Family ass 1})]}\\
    \leq & (1+\alpha)\sup_{i\geq 1}\sup_{m\leq k}\bigg|\left(\pdv{\boldsymbol\lambda_i}{\theta_j}\right)_m\bigg|\left(\sum_{m=1}^k\bigg[|(\boldsymbol C_\alpha(\boldsymbol\lambda_i))_m|\bigg]+kM^\alpha \|\boldsymbol u_\lambda(X,\boldsymbol\lambda_i)\|_\infty\right). 
\end{align*}
Now, observe that, by (\ref{Single Family ass 1}) we have, for all $m\leq k$, $|(\boldsymbol C_\alpha(\boldsymbol\lambda_i))_m|$ a continuous function of $\lambda_i\in \Lambda$. 
Also, $\Lambda$ being a compact space we have  $|(\boldsymbol C_\alpha(\boldsymbol\lambda_i))_m|$ bounded for all $m$. Hence, there exists a $B>0$ such that
\[
\sum_{m=1}^k|(\boldsymbol C_\alpha(\boldsymbol\lambda_i))_m|\leq B.
\]
Thus, we have 
\[
|\boldsymbol\nabla_j V_i(X;\boldsymbol\theta,\boldsymbol{t})|\leq  (1+\alpha)\sup_{i\geq 1}\sup_{m\leq k}\bigg|\left(\pdv{\boldsymbol\lambda_i}{\theta_j}\right)_m\bigg|\left[B+kM^\alpha \|\boldsymbol u_\lambda(X,\boldsymbol\lambda_i)\|_\infty\right].
\]
Further, note the following implication
\begin{align*}
    (1+\alpha)\sup_{i\geq 1}\sup_{m\leq k}&\bigg|\left(\pdv{\boldsymbol\lambda_i}{\theta_j}\right)_m\bigg|\left[B+kM^\alpha \|\boldsymbol u_\lambda(X,\boldsymbol\lambda_i)\|_\infty\right]>N.
    \\
    \implies \|\boldsymbol u_\lambda(X,\boldsymbol\lambda_i)\|_\infty&>\frac{N}{kM^\alpha(1+\alpha)\sup_{i\geq 1}\sup_{m\leq k}\bigg|\left(\pdv{\boldsymbol\lambda_i}{\theta_j}\right)_m\bigg|}-\frac{B}{kM^\alpha}:=N_0.
\end{align*}
Now, since by( \ref{Single Family ass 2}) the terms in denominator of $N_0$ are finite, we have
\[
N\to \infty \implies N_0\to \infty
\]
Then, using Lemma (\ref{single increment family lemma}), we get
\begin{align*}
    \frac{1}{n}\sum_{i=1}^n & E_{\boldsymbol\lambda_i}\bigg[|\boldsymbol\nabla_j V_i(X;\boldsymbol\theta,\boldsymbol{t})|I(|\boldsymbol\nabla_j V_i(X;\boldsymbol\theta,\boldsymbol{t})|>N)\bigg]\\
    \leq \frac{1}{n}\sum_{i=1}^n & E_{\boldsymbol\lambda_i}\bigg[(1+\alpha)\sup_{i\geq 1}\sup_{m\leq k}\bigg|\left(\pdv{\boldsymbol\lambda_i}{\theta_j}\right)_m\bigg|\left[B+kM^\alpha \|\boldsymbol u_\lambda(X,\boldsymbol\lambda_i)\|_\infty\right]\\
    I&\left((1+\alpha)\sup_{i\geq 1}\sup_{m\leq k}\bigg|\left(\pdv{\boldsymbol\lambda_i}{\theta_j}\right)_m\bigg|\left[B+kM^\alpha \|\boldsymbol u_\lambda(X,\boldsymbol\lambda_i)\|_\infty\right]>N\right)\bigg]\\
    =&(1+\alpha)\sup_{i\geq 1}\sup_{m\leq k}\bigg|\left(\pdv{\boldsymbol\lambda_i}{\theta_j}\right)_m\bigg|B\left(\frac{1}{n}\sum_{i=1}^n P_{\boldsymbol\lambda_i}(\|\boldsymbol u_\lambda(X,\boldsymbol\lambda_i)\|_\infty>N_0)\right)\\
    +&(1+\alpha)\sup_{i\geq 1}\sup_{m\leq k}\bigg|\left(\pdv{\boldsymbol\lambda_i}{\theta_j}\right)_m\bigg|kM^\alpha\left(\frac{1}{n}\sum_{i=1}^n E_{\lambda_i}\left(\|\boldsymbol u_\lambda(X,\boldsymbol\lambda_i)\|_\infty I(\|\boldsymbol u_\lambda(X,\boldsymbol\lambda_i)\|_\infty>N_0)\right)\right).
\end{align*}
Now, finally by Assumptions (\ref{Single Family ass 2})--(\ref{Single Family ass 3}) we can see that first condition of (\ref{Indep ass 6}) holds true.

\textit{Second condition of (\ref{Indep ass 6}):} It will follow in the same way as the first condition did, and thus, skipped here for brevity.

\textit{Condition of (I3):}
To begin with, we will state two elementary results on norms we are gonna use. 
Firstly, for any given matrix $\boldsymbol{M}$ and vector $\boldsymbol{x}$, we have
\begin{equation}\label{Result 1}
 \|\boldsymbol{M}\boldsymbol{x}\|\leq \|\boldsymbol{M}\|\|\boldsymbol{x}\|,   
\end{equation}
which directly follows from the definition. 
Our second result is as follows. For any two given vectors $\boldsymbol{a}, \boldsymbol{b}\in \mathbb{R}^p$,we have
\begin{equation}\label{Result 2}
 \|\boldsymbol{a}-\boldsymbol{b}\|^2\leq 2(\||\boldsymbol{a}|\|^2+\||\boldsymbol{b}|\|^2).   
\end{equation}
where $|\boldsymbol{a}|\in \mathbb{R}^p$ is defined as $|\boldsymbol{a}|_i=|a_i|$ for $i\leq p$. 
This second result (\ref{Result 2}) can be proved easily as follows:
\begin{align*}
    \|\boldsymbol{a}-\boldsymbol{b}\|^2= \sum_{i=1}^p (a_i-b_i)^2
     \leq  \sum_{i=1}^p 2(|a_i|^2+|b_i|^2)
     =2(\||\boldsymbol{a}|\|^2+\||\boldsymbol{b}|\|^2).
\end{align*}

Now, by Result (\ref{Result 1}) and Assumption (\ref{Single Family ass 2}), there exists $B>0$ such that
\[
\|\boldsymbol\Omega_n^{-1/2}\boldsymbol\nabla V_i(Y;\boldsymbol\theta,\boldsymbol{t})\|\leq \|\boldsymbol\Omega_n^{-1/2}\|\|\boldsymbol\nabla V_i(Y;\boldsymbol\theta,\boldsymbol{t})\|\leq B \|\boldsymbol\nabla V_i(Y;\boldsymbol\theta,\boldsymbol{t})\|.
\]
Recall, we have for single family of increment distributions and hence
\[
\boldsymbol\nabla_j V_i(X;\boldsymbol\theta,\boldsymbol{t})=(1+\alpha)\left(\pdv{\boldsymbol\lambda_i}{\theta_j}\right)^T[\boldsymbol C_\alpha(\boldsymbol\lambda_i)-f^\alpha(X,\boldsymbol\lambda_i)\boldsymbol u_\lambda(X,\boldsymbol\lambda_i)].
\]
In a more compact matrix notations, we can write 
\[
\boldsymbol\nabla V_i(X;\boldsymbol\theta,\boldsymbol{t})=(1+\alpha)\boldsymbol\Lambda_i^T[\boldsymbol C_\alpha(\boldsymbol\lambda_i)-f^\alpha(X,\boldsymbol\lambda_i)\boldsymbol u_\lambda(X,\boldsymbol\lambda_i)].
\]
Now, note that
\begin{align*}
    \|\boldsymbol\nabla V_i(X;\boldsymbol\theta,\boldsymbol{t})\|
    \leq (1+\alpha)\|\boldsymbol\Lambda_i^T\|\times \|\boldsymbol C_\alpha(\boldsymbol\lambda_i)-f^\alpha(X,\boldsymbol\lambda_i)\boldsymbol u_\lambda(X,\boldsymbol\lambda_i)\|.
\end{align*}
Then, by (\ref{Single Family ass 2}), one can show that there exists $K$ satisfying 
$\|\boldsymbol\Lambda_i^T\|\leq K$ for all $i$.

Further, using (\ref{Single Family ass 1}) again along with Result (\ref{Result 2}), we get 
\[
\|\boldsymbol C_\alpha(\boldsymbol\lambda_i)-f^\alpha(X,\boldsymbol\lambda_i)\boldsymbol u_\lambda(X,\boldsymbol\lambda_i)\|^2\leq  2\|\boldsymbol C_\alpha(\boldsymbol\lambda_i)\|^2+2M^{2\alpha}\|\boldsymbol u_\lambda(X,\boldsymbol\lambda_i)\|^2.
\]
Next, using (\ref{Single Family ass 1}) again, $\|\boldsymbol C_\alpha(\boldsymbol\lambda_i)\|^2$ is a continuous function of $\boldsymbol\lambda_i$, and hence, is bounded by some $L>0$. 
Thus, using Lemma \ref{single increment family lemma}, we get 
\begin{align*}
   &\sum_{i=1}^n E_{\boldsymbol\lambda_i}[\|\boldsymbol\Omega_n(\boldsymbol{t})^{-1/2}\boldsymbol\nabla V_i(Y;\boldsymbol\theta,\boldsymbol{t})\|^2I(\|\boldsymbol\Omega_n(\boldsymbol{t})^{-1/2}\boldsymbol\nabla V_i(Y;\boldsymbol\theta,\boldsymbol{t})\|>\epsilon\sqrt{n})]\\
   \leq &\sum_{i=1}^n 2(B(1+\alpha)K)^2 E_{\boldsymbol\lambda_i}\left[\left(L+M^{2\alpha}\|\boldsymbol u_\lambda(X,\boldsymbol\lambda_i)\|^2\right)I\left(L+M^{2\alpha}\|\boldsymbol u_\lambda(X,\boldsymbol\lambda_i)\|^2>\frac{\epsilon^2n}{2(B(1+\alpha)K)^2}\right)\right].
\end{align*}
Finally, proceeding like what we showed above for (\ref{Indep ass 6}), similar results for (\ref{Indep Ass 7}) will follow too.
\hfill{$\square$}

\subsection{Proof of Corollary \ref{Poisson family theorem}}


First observe that, here $\ell= 1$. Also, just by elementary calculus one can verify for $f(X,\lambda_i)= e^{-\lambda_i}\lambda_i^X/X!$ that
\begin{align*}
    u_\lambda(X,\lambda_i)= \frac{X-\lambda_i}{\lambda_i},~~~~
\nabla_\lambda u_\lambda(X,\lambda_i)= -\frac{X}{\lambda_i^2}~~~~\mbox{ and }~~~
u_\lambda(X,\lambda_i)u_\lambda(X,\lambda_i)^T= \frac{(X-\lambda_i)^2}{\lambda_i^2}.
\end{align*}
Thus, following our notation from Section \ref{SEC:single family}, we have
\begin{align*}
C_{\alpha}(\lambda_i)=\sum_{k\in \mathcal{N}}&\frac{e^{-\lambda_i(1+\alpha)}\lambda_i^{k(1+\alpha)}(k-\lambda_i)}{(k!)^{1+\alpha}\lambda_i},\\
C_{\alpha}^{(1)}(\lambda_i)=\sum_{k\in \mathcal{N}}-\frac{e^{-\lambda_i(1+\alpha)}\lambda_i^{k(1+\alpha)}k}{(k!)^{1+\alpha}\lambda_i^2}&~~~\mbox{and}~~~
C_{\alpha}^{(2)}(\lambda_i)=\sum_{k\in \mathcal{N}}\frac{e^{-\lambda_i(1+\alpha)}\lambda_i^{k(1+\alpha)}(k-\lambda_i)^2}{(k!)^{1+\alpha}\lambda_i^2}.\\
\end{align*}
Accordingly, we get
\[
\Psi_n(\boldsymbol{t})=\sum_{i=1}^n \frac{(1+\alpha)}{n}\boldsymbol\Lambda_i^TC_{\alpha,i}^{(2)}\boldsymbol\Lambda_i,
~~~
\mbox{ and }
~~~
\quad \Omega_n(\boldsymbol{t})= \frac{1}{n}\sum_{i=1}^n (1+\alpha)^2\boldsymbol\Lambda_i^T \left[C_{2\alpha,i}^{(2)}-C_{\alpha,i}^2\right]\boldsymbol\Lambda_i.
\]
Further, in terms of notation of (\ref{Single Family ass 3}), we have 
With our defined notations, we also have
\begin{align*}
W_{2i}= \frac{|X-\lambda_i|}{\lambda_i}+\frac{|X|}{\lambda_i^2}+ \frac{(X-\lambda_i)^2}{\lambda_i^2},
\end{align*}
and additionally define $W_{1i}=W_{3i}= {|X-\lambda_i|}/{\lambda_i}$.
To study the tail bounds of these quantities, let us note the following lemma on Poisson Distribution; its proof is given in Appendix C.4.

\begin{lemma}\label{poisson process lemma}
	Suppose $X\sim Poi(\lambda)$. Then following holds for large enough $k\in \mathbb{N}$
	\begin{align*}
	P_\lambda(|X|>k)= O\left(\frac{1}{\sqrt{k}}\right),~~~
	E_\lambda(|X|I(|X|>k)= O\left(\frac{\lambda}{\sqrt{k}}\right)~~~\mbox{ and }~~~
	E_\lambda(X^2I(|X|>k) = O\left(\frac{\lambda^2}{\sqrt{k}}\right).
	\end{align*}
\end{lemma}

Now, in order to show that (\ref{Single Family ass 3}) follows directly from  (\ref{Single Family ass 1})--(\ref{Single Family ass 2}) under the Poisson process model, 
we observe that, for large $N$ and any $i$, using above Lemma \ref{poisson process lemma} we have
\begin{align*}
P_{\lambda_i}(W_{1i}>N)= P_{\lambda_i}(X-\lambda_i>N\lambda_i)
= O\left(\frac{1}{(\sqrt{(N+1)\lambda_i}}\right)
= O\left(\frac{1}{\sqrt{N\lambda_i}}\right)
=  O\left(\frac{1}{\sqrt{N}}\right), 
\end{align*}
where the last step follows using (\ref{Single Family ass 1}).
Thus, one can note that, for all $n$, 
\[
\frac{1}{n}\sum_{i=1}^n P_{\lambda_i}(W_{1i}>N)= O\left(\frac{1}{\sqrt{N}}\right).
\]
Now, similarly observe that 
\begin{align*}
E_{\lambda_i}[W_{1i}I(W_{1i}>N)]& = E_{\lambda_i}[XI(X>(N+1)\lambda_i)]-\lambda_i P_{\lambda_i}(X>(N+1)\lambda_i)\\
& \leq E_{\lambda_i}[XI(X>(N+1)\lambda_i)]\\
& = O\left(\frac{\lambda_i}{\sqrt{(N+1)\lambda_i}}\right)= O\left(\sqrt{\frac{\lambda_i}{N}}\right)= O\left(\frac{1}{\sqrt{N}}\right) \quad (\text{Using (SF1)}).
\end{align*}
Hence, the third condition of (\ref{Single Family ass 3}) follows directly for Poisson Process if (\ref{Single Family ass 1}) holds. 
Also, the second condition of (\ref{Single Family ass 3}) will follow similarly as above assuming (\ref{Single Family ass 1})--(\ref{Single Family ass 2}).

Lastly, observe the following set inequality:
\begin{align*}
\{W_{2i}>N\}\subseteq & \left\{\frac{|X-\lambda_i|}{\lambda_i}>N/3\right\}\cup\left\{\frac{|X|}{\lambda_i^2}>N/3\right\}\cup\left\{\frac{(X-\lambda_i)^2}{\lambda_i^2}>N/3\right\}\\
\subseteq & \left\{ X>(N/3+1)\lambda_i\right\}\cup\left\{X>N\lambda_i^2/3\right\}\cup\left\{X>(\sqrt{N/3}+1)\lambda_i\right\}.
\end{align*}
Thus, we have
\begin{align*}
P(V_{2i}>N)& \leq P(X>(N/3+1)\lambda_i)+P(X>N\lambda_i^2/3)+P(X>(\sqrt{N/3}+1)\lambda_i)\\
& = O\left(\frac{1}{\sqrt{N}}\right)+O\left(\frac{1}{\sqrt{N}}\right)+O\left(\frac{1}{N^{1/4}}\right) \quad (\text{Proceeding as before})\\
& = O\left(\frac{1}{\sqrt{N}}\right).
\end{align*}
Hence, first condition of (\ref{Single Family ass 3}) can be proved following a similar path.
\hfill{$\square$}

\subsection{Proof of Corollary \ref{location scale family theorem}}

Suppose that $\boldsymbol\mu_i$ is the $i$-th location parameter and $\sigma_i$ is the $i$-th scale parameter and the corresponding location-scale family is 
\[
f(X;\boldsymbol\mu_i,\sigma_i)=\frac{1}{\sigma_i}f\left(\frac{X-\boldsymbol\mu_i}{\sigma_i}\right).
\]
Thus, one can check that 
\begin{align*}
\boldsymbol u_\lambda(X,\boldsymbol\lambda_i)= & \left[-\frac{1}{\sigma_i}\frac{f'\left(\frac{X-\boldsymbol\mu_i}{\sigma_i}\right)}{f\left(\frac{X-\boldsymbol\mu_i}{\sigma_i}\right)}\nabla_\lambda\boldsymbol{\mu}_i, 
-\frac{1}{\sigma_i}\left(1+\frac{f'\left(\frac{X-\boldsymbol\mu_i}{\sigma_i}\right)^T\left(\frac{X-\boldsymbol\mu_i}{\sigma_i}\right)}{f\left(\frac{X-\boldsymbol\mu_i}{\sigma_i}\right)}\right)\nabla_\lambda\sigma_i\right]
= : \frac{1}{\sigma_i}G\left(\frac{X-\boldsymbol\mu_i}{\sigma_i}\right),
\end{align*}
for some suitable function $G$, where $f'(\cdot)$ is the derivative of $f(\cdot)$. 
Similarly, one can find suitable functions $G^{(1)}$ and $G^{(2)}$ such that
$$
\nabla_{\boldsymbol{\lambda}}\boldsymbol u_\lambda(X,\boldsymbol\lambda_i)=:\frac{1}{\sigma_i^2}G^{(1)}\left(\frac{X-\mu_i}{\sigma_i}\right),~~~~\text{and}~~~~
\boldsymbol u_\lambda(Y,\boldsymbol\lambda_i)\boldsymbol u_\lambda(Y,\boldsymbol\lambda_i)^T=:\frac{1}{\sigma_i^2}G^{(2)}\left(\frac{X-\mu_i}{\sigma_i}\right).
$$
Hence, we can do the following simplifications
\begin{align*}
\boldsymbol C_{\alpha}(\boldsymbol\lambda_i)=&E_{\boldsymbol\mu_i,\sigma_i}\left(\frac{1}{\sigma_i^{1+\alpha}}f^\alpha\left(\frac{X-\boldsymbol\mu_i}{\sigma_i}\right)G\left(\frac{X-\boldsymbol\mu_i}{\sigma_i}\right)\right)=\frac{1}{\sigma_i^{1+\alpha}}E_{\boldsymbol 0,1}\left(f^\alpha(X)G(X)\right),\\
\boldsymbol C^{(1)}_{\alpha}(\boldsymbol\lambda_i)=&E_{\boldsymbol\mu_i,\sigma_i}\left(\frac{1}{\sigma_i^{2+\alpha}}f^\alpha\left(\frac{X-\boldsymbol\mu_i}{\sigma_i}\right)G^{(1)}\left(\frac{X-\boldsymbol\mu_i}{\sigma_i}\right)\right)=\frac{1}{\sigma_i^{2+\alpha}}E_{\boldsymbol 0,1}\left(f^\alpha(X)G^{(1)}(X)\right),\\
\boldsymbol C^{(2)}_{\alpha}(\boldsymbol\lambda_i)=&E_{\boldsymbol\mu_i,\sigma_i}\left(\frac{1}{\sigma_i^{2+\alpha}}f^\alpha\left(\frac{X-\boldsymbol\mu_i}{\sigma_i}\right)G^{(2)}\left(\frac{X-\boldsymbol\mu_i}{\sigma_i}\right)\right)=\frac{1}{\sigma_i^{2+\alpha}}E_{\boldsymbol 0,1}\left(f^\alpha(X)G^{(2)}(X)\right).
\end{align*}
Using these simplified results, one can easily observe that 
\[
\boldsymbol \Psi_n(\boldsymbol{t})=\sum_{i=1}^n \frac{(1+\alpha)}{n\sigma_i^{2+\alpha}}\boldsymbol\Lambda_i^T\boldsymbol C_\alpha^{(2)}(\boldsymbol 0,1)\boldsymbol\Lambda_i,
\]
and
\[
\quad \boldsymbol\Omega_n(\boldsymbol{t})= \frac{1}{n}\sum_{i=1}^n \frac{(1+\alpha)^2}{\sigma_i^{2+2\alpha}}\boldsymbol\Lambda_i^T \left[\boldsymbol C_{2\alpha}^{(2)}(\boldsymbol 0,1)-\boldsymbol C_{\alpha}(\boldsymbol 0,1)\boldsymbol C_\alpha(\boldsymbol 0,1)^T\right]\boldsymbol\Lambda_i.
\]

But, in terms of the notation from (\ref{Single Family ass 3}), we have
$$
W_{2i}(Y)= \|\boldsymbol u_\lambda(Y,\boldsymbol\lambda_i)\|_\infty
+  \|\nabla_{\boldsymbol{\lambda}}\boldsymbol u_\lambda(Y,\boldsymbol\lambda_i)\|_\infty,
+ \|\boldsymbol u_\lambda(Y,\boldsymbol\lambda_i)\boldsymbol u_\lambda(Y,\boldsymbol\lambda_i)^T\|_\infty.
$$
For simplicity of notation, we further define
    $$ 
W_{1i}(Y):= \|\boldsymbol u_\lambda(Y,\boldsymbol\lambda_i)\|_\infty,~~~
W_{3i}(Y):= \|\boldsymbol u_\lambda(Y,\boldsymbol\lambda_i)\|_2.
$$ 
Then, following the discussion above, we can observe that there exists a suitable non-negative functions $G_1$ and $G_3$ independent of $i$ such that
\[
W_{1i}(Y)=\frac{1}{\sigma_i}G_1\left(\frac{Y-\boldsymbol\mu_i}{\sigma_i}\right)~~~\text{and}~~~
W_{3i}(Y)=\frac{1}{\sigma_i}G_3\left(\frac{Y-\boldsymbol\mu_i}{\sigma_i}\right).
\]
and further we can find some non-negative functions $G_{21}$ and $G_{22}$ independent of $i$, such that
\[
W_{2i}=\frac{1}{\sigma_i}G_{21}\left(\frac{X-\boldsymbol\mu_i}{\sigma_i}\right)+\frac{1}{\sigma_i^2}G_{22}\left(\frac{X-\boldsymbol\mu_i}{\sigma_i}\right).
\]

Now, using (\ref{Single Family ass 1}) we also have $\frac{1}{\sigma_i}$ bounded for all $i$. Then we have, for some $B>0$, 
\begin{align*}
W_{1i}\leq B G_1\left(\frac{X-\boldsymbol\mu_i}{\sigma_i}\right) \quad \text{and} \quad W_{2i}\leq & B G_{21}\left(\frac{X-\boldsymbol\mu_i}{\sigma_i}\right)+B^2 G_{22}\left(\frac{X-\boldsymbol\mu_i}{\sigma_i}\right)\\
\implies W_{2i}\leq & G_2\left(\frac{X-\boldsymbol\mu_i}{\sigma_i}\right) \quad \text{ for some function $G_2$ }.
\end{align*}
Thus, we have the following for all $i$:
\begin{align*}
P_{\lambda_i}[W_{1i}>N]\leq & P_{\boldsymbol\mu_i,\sigma_i}\left(B G_1\left(\frac{X-\boldsymbol\mu_i}{\sigma_i}\right)>N\right) \\
= & P_{\boldsymbol{0},1}(BG_1(X)>N).
\end{align*}
Hence, we have
\[
\lim_{N\to \infty} \sup_{n>1} \frac{1}{n}\sum_{i=1}^n P_{\boldsymbol\lambda_i}[W_{1i}>N]\leq \lim_{N\to \infty} P_{\boldsymbol 0,1}(G_1(X)>N/B) = 0.
\]
Hence, the first part of the third condition of (\ref{Single Family ass 3}) holds directly for location scale family. The second condition also follows directly, if we first observe
\[
\lim_{N\to \infty} \sup_{n>1} \frac{1}{n}\sum_{i=1}^n E_{\boldsymbol\lambda_i}[W_{1i}I(W_{1i}>N)]\leq \lim_{N\to\infty}E_{\boldsymbol 0,1}[BG_1(X)I(BG_1(X)>N)] = 0.
\]
In th above, the first inequality follows from Lemma \ref{single increment family lemma} and the second equality follows by an application of the dominated convergence theorem (DCT), 
assuming that $\boldsymbol u_\lambda(X,\boldsymbol \lambda_i)$ is integrable. 
Hence, the third condition of (\ref{Single Family ass 3}) follows. 

Similarly we can also show similar tail behavior for $W_{2i}$ and $W_{3i}$ as required in the first and second conditions of (\ref{Single Family ass 3}).
\hfill{$\square$}

\subsection{Proof of Corollary \ref{DB family theorem}}
\label{proof of DB family theorem}

Recall that the drifted Brownian motion falls into category of location-scale family of increment distribution, 
and thus, the result follows for this IIP readily from Corollary \ref{location scale family theorem}.
So, we only need to simplify the expressions for $\boldsymbol\Psi_n$ and $\boldsymbol\Omega_n$ as follows.

Here, as per the notation for the location scale family of increment distributions, we have $f(\cdot)$ to be the  standard normal density, i.e.,
\[
f(x)=\frac{1}{\sqrt{2\pi}}e^{-x^2/2}.
\]
Hence, with the notation from the proof of Corollary \ref{location scale family theorem}, one can check that
\begin{align*}
G(x) = \left(x,x^2-1\right),
~~~~
G^{(1)}(x) = \begin{bmatrix}
-1 & -2x\\
-2x & 1-3x^2
\end{bmatrix}~~~ &\mbox{ and } ~~~
G^{(2)}(x)  = \begin{bmatrix}
x^2 & x^3-x\\
x^3-x & (x^2-1)^2
\end{bmatrix}.
\end{align*}
Note, in this case,
$f^{1+\alpha}(x)$ = $(2\pi)^{-\frac{\alpha}{2}}\frac{1}{\sqrt{1+\alpha}}$ times the pdf of Normal distribution with mean 0 variance $\frac{1}{1+\alpha}$.
Thus, one can compute that
\begin{align*}
\boldsymbol C_{\alpha}(\boldsymbol\lambda_i)  = \frac{(2\pi)^{-\frac{\alpha}{2}}}{\sqrt{1+\alpha}\sigma_i^{1+\alpha}}&\times \left(0,\frac{1}{1+\alpha}-1\right),
\\
\boldsymbol C_{\alpha}^{(1)}(\boldsymbol\lambda_i)  = \frac{(2\pi)^{-\frac{\alpha}{2}}}{\sqrt{1+\alpha}\sigma_i^{2+\alpha}}\times \begin{bmatrix}
-1 & 0\\
0 & 1-\frac{3}{1+\alpha}
\end{bmatrix},
~~~~\mbox{ and }&~~~~
\boldsymbol C_{\alpha}^{(2)}(\boldsymbol\lambda_i)  = \frac{(2\pi)^{-\frac{\alpha}{2}}}{\sqrt{1+\alpha}\sigma_i^{2+\alpha}}\times \begin{bmatrix}
\frac{1}{1+\alpha} & 0\\
0 & \frac{3}{(1+\alpha)^2}-\frac{2}{1+\alpha}+1
\end{bmatrix},
\end{align*}
and thus, it is easy to verify that
\[
\boldsymbol\Psi_n(\boldsymbol{t})=\sum_{i=1}^n \frac{(2\pi)^{-\frac{\alpha}{2}}(1+\alpha)^{1/2}}{n\sigma_i^{2+\alpha}}\boldsymbol\Lambda_i^T \begin{bmatrix}
\frac{1}{1+\alpha} & 0\\
0 & \frac{3}{(1+\alpha)^2}-\frac{2}{1+\alpha}+1
\end{bmatrix}\boldsymbol\Lambda_i,
\]
and
\[
\quad \boldsymbol\Omega_n(\boldsymbol{t})= \frac{1}{n}\sum_{i=1}^n \frac{(1+\alpha)^{2}(2\pi)^{-\alpha}}{\sigma_i^{2+2\alpha}\sqrt{1+2\alpha}}\boldsymbol\Lambda_i^T \begin{bmatrix}
\frac{1}{1+2\alpha} & 0\\
0 & \frac{3}{(1+2\alpha)^{2}}-\frac{2}{1+2\alpha}+1-\frac{\sqrt{1+2\alpha}}{1+\alpha}\left(\frac{1}{1+\alpha}-1\right)^2
\end{bmatrix}\boldsymbol\Lambda_i.
\]
\hfill{$\square$}

\section{Proofs of the Lemmas}
\label{lemma proofs}

\subsection{Proof of Lemma \ref{markov process wlln lemma}}

Fix $\epsilon>0$ and $n\in \mathrm{N}$. 
Then, using Chebychev's Inequality, we have
$$P\left[\bigg|\frac{Sn-E(S_n)}{n}\bigg|>\epsilon\right]\leq \frac{Var(S_n/n)}{\epsilon^2}.$$
Now, first observe that
\begin{align*}
    Var\left(\frac{S_n}{n}\right)=\frac{Var(S_n)}{n^2}
    \leq \sigma^2 \frac{\sum_i^n\sum_j^n |Cov(X_i,X_j)|}{n^2}\leq \sigma^2 \frac{\sum_i^n\sum_j^n c^{|i-j|} }{n^2}.
\end{align*}
But, we can do the following calculations:
\begin{align*}
    \sum_{i=1}^n\sum_{j=1}^n c^{|i-j|}=\sum_{i=1}^n 1 +2\sum_{i=1}^n\sum_{j=1}^{i-1}c^{|i-j|} =&n+2\sum_{i=1}^n\sum_{j=1}^{i-1}c^{i-j}\\
    =&n+2c\sum_{i=1}^n \frac{1-c^{i-1}}{1-c}\\
    \leq &n+\frac{2c}{1-c}\sum_{i=1}^n 1=n\frac{1+c}{1-c}.
\end{align*}
Thus, finally we have
\begin{align*}
    \frac{Var(S_n/n)}{\epsilon^2}\leq & \frac{n\left(\frac{1+c}{1-c}\right)}{n^2\epsilon^2}\to 0 \text{ as } n\to \infty.
\end{align*}
Since $\epsilon$ is arbitrary, we can conclude that
\[
\frac{S_n-E(S_n)}{n} \overset{P}{\to}0.
\]
\hfill{$\square$}

\subsection{Proof of Lemma \ref{m dependent wlln lemma}}
Fix $\epsilon>0$ and $n\in \mathbb{N}$. 
Using Chebychev's Inequality, we have
$$P\left[\bigg|\frac{Sn-E(S_n)}{n}\bigg|>\epsilon\right]\leq \frac{Var(S_n/n)}{\epsilon^2}.$$
Now, first observe that, for all $i,j$, by Cauchy-Schwartz inequality, we get
\[
|Cov(X_i,X_j)|\leq \sqrt{Var(X_i)Var(X_j)}=\sigma^2.
\]
Therefore, 
\begin{align*}
    Var\left(\frac{S_n}{n}\right)=\frac{Var(S_n)}{n^2}
    \leq &\frac{\sum_i^n\sum_j^n |Cov(X_i,X_j)|}{n^2}\\
    \leq &\frac{\sum_i^n\sum_{j=max(i-m,0)}^{min(n,i+m)} |Cov(X_i,X_j)|}{n^2}
    \leq \frac{\sum_{i=1}^n2m\sigma^2}{n^2}
    \leq \frac{nm\sigma^2}{n^2}\to 0,
\end{align*}
and the result follows.
\hfill{$\square$}

\subsection{Proof of Lemma \ref{m dependent lemma 2}}

To show the result, let us first observe that 
\begin{align*}
    \frac{Var(S_n)}{n}=\frac{Var(S_n)}{n}
    \leq \frac{\sum_i^n\sum_j^n |Cov(X_i,X_j)|}{n}
    = &\frac{\sum_i^n\sum_{j=max(i-m,0)}^{min(n,i+m)} |Cov(X_i,X_j)|}{n}\\
    \underset{\text{By stationarity}}{\leq} & \frac{\sum_{i=1}^n\sum_{j=i-m}^{i+m}|Cov(X_1,X_j)|}{n}=\frac{\sum_{i=1}^nA}{n}=A.
\end{align*} 
Similarly, also observe that
\begin{align*}
    \frac{Var(S_n)}{n}
    \geq  \frac{\sum_{i=m+1}^{n-m}\sum_j^n |Cov(X_i,X_j)|}{n}
    \underset{\text{By stationarity}}{\geq} & \frac{\sum_{i=m+1}^{n-m}A}{n}\geq  \frac{(n-2m)A}{n}.
\end{align*}
Hence, by Sandwich theorem, we have
$
\frac{Var(S_n)}{n}\to A.
$
\hfill{$\square$}

\subsection{Proof of Lemma \ref{poisson process lemma}}

Observe firstly that, for $k\in \mathbb{N}$, we have
\[
P_\lambda(X=k)=e^{-\lambda}\frac{\lambda^k}{k!}.
\]
Thus, we can see that
\begin{align*}
    P_\lambda(|X|>k)=&\sum_{i=k+1}^\infty e^{-\lambda}\frac{\lambda^i}{i!}\\
    \approx & \sum_{i=k+1}^\infty e^{-\lambda} \frac{\lambda^ie^i}{(\sqrt{2\pi i})i^i} & \left(\text{Stirling Approximation } i!\approx \sqrt{2\pi i}(i/e)^i\right) \\
    =&\sum_{i=k+1}^\infty e^{-\lambda}\left(\frac{\lambda e}{i}\right)^i\frac{1}{\sqrt{2\pi i}}\\
    \leq &\sum_{i=k+1}^\infty e^{-\lambda}\left(\frac{\lambda e}{k}\right)^i\frac{1}{\sqrt{2\pi k}} & (i>k)\\
    =& e^{-\lambda} \frac{1}{\sqrt{2\pi k}}\frac{1}{1-\frac{\lambda e}{k}}\left(\frac{\lambda e}{k}\right)^k & \left(\text{$k$ is big, $\left(\frac{\lambda e}{k}\right)<1$}\right) \\
    \leq & 1\times \frac{1}{\sqrt{k}}\times \frac{k}{k-\lambda e}\times 1 \\
    =&O\left(\frac{1}{\sqrt{k}}\right) & \left(\frac{k}{k-\lambda e}\approx 1\right).
\end{align*}
Next, similarly we can observe that
\begin{align*}
    E_\lambda(|X|I(|X|>k))=&\sum_{i=k+1}^\infty ie^{-\lambda}\frac{\lambda^i}{i!}=\sum_{i=k}^\infty \lambda e^{-\lambda}\frac{\lambda^i}{i!}
    =\lambda P_\lambda(X>k-1)
    =O\left(\frac{\lambda}{\sqrt{k}}\right).
\end{align*}
Similarly, we also can show
\[
E_\lambda(X^2I(|X|>k))=O\left(\frac{\lambda^2}{\sqrt{k}}\right).
\]
\hfill{$\square$}

\biblio 

\end{appendices}

\bibliographystyle{apaeng} 

\renewcommand\refname{References} 
{
	\addcontentsline{toc}{section}{References} 
	\bibliography{References.bib} 
}

\end{document}